\documentclass[fleqn,usenatbib,useAMS]{mnras}

\usepackage{graphicx}	
\usepackage{amsmath}	
\usepackage{amssymb}	
\usepackage{multicol}        
\usepackage{bm}		
\usepackage{pdflscape}	

\usepackage{siunitx}
\usepackage{comment}
\usepackage{xspace}
\usepackage[dvipsnames]{xcolor}
\usepackage{cite}

\usepackage{multirow}
\usepackage{array}
\newcolumntype{x}[1]{>{\centering\arraybackslash\hspace{0pt}}p{#1}}

\usepackage[T1]{fontenc}

\DeclareRobustCommand{\VAN}[3]{#2}
\let\VANthebibliography\thebibliography
\def\thebibliography{\DeclareRobustCommand{\VAN}[3]{##3}\VANthebibliography}

\newcommand{\ProSpect}{\textsc{ProSpect}\xspace}
\newcommand{\ProFound}{\textsc{ProFound}\xspace}
\newcommand{\Hyperfit}{\textsc{Hyperfit}\xspace}

\newcommand{\LTIR}{$L_\mathrm{TIR}$\xspace}
\newcommand{\Lradio}{$L_\mathrm{1.4\,GHz}$\xspace}
\newcommand{\Mstar}{$M_{\star}$\xspace}
\newcommand{\SFRburst}{$\mathrm{SFR_{burst}}$\xspace}
\newcommand{\SFRt}{SFR(t)\xspace}
\newcommand{\DeltaSFRGyr}{$\Delta \mathrm{SFR_{1\,Gyr}}$\xspace}
\newcommand{\DeltaSFRMyr}{$\Delta \mathrm{SFR_{200\,Myr}}$\xspace}

\newcommand{\qIR}{$q_\mathrm{IR}$\xspace}
\newcommand{\qTIR}{$q_\mathrm{TIR}$\xspace}
\newcommand{\qFIR}{$q_\mathrm{FIR}$\xspace}
\newcommand{\aveqTIR}{$\langle q_\mathrm{TIR}\rangle$\xspace}
\newcommand{\qSFR}{$q_\mathrm{SFR}$\xspace}

\newcommand{\HI}{H$\,\textsc{i}$\xspace}    

\newcommand{\fAGN}{$f_\mathrm{AGN}$\xspace}  

\newcommand{\muJybeam}{$\mu$Jy\,beam$^{-1}$\xspace}

\newcommand{\WperHz}{W\,Hz$^{-1}$\xspace}
\newcommand{\Msun}{M$_{\sun}$\xspace}

\newcommand{\MSunperyear}{$\mathrm{M_{\sun}\,yr^{-1}}$\xspace}

\newcommand{\microns}{$\mathrm{\mu m}$\xspace}
\newcommand{\degsq}{$\mathrm{deg}^{2}$\xspace}

\newcommand{\xtimes}{{\mkern-2mu\times\mkern-2mu}}
\newcommand{\DEVILSxMIGHTEE}{DEVILS$\times$MIGHTEE\xspace}
\newcommand{\GAMAxDINGO}{GAMA$\times$DINGO\xspace}

\title[Impact of Timescales on SFR-L1.4GHz Relation]{DEVILS/MIGHTEE/GAMA/DINGO: The Impact of SFR Timescales on the SFR-Radio Luminosity Correlation}

\author[R. H. W. Cook et al.]{Robin H. W. Cook,$^{1}$\thanks{E-mail: robin.cook@uwa.edu.au}
Luke J. M. Davies,$^{1}$
Jonghwan Rhee,$^{1,2}$
Catherine L. Hale,$^{3}$
Sabine Bellstedt,$^{1}$\newauthor
Jessica E. Thorne,$^{1}$
Ivan Delvecchio,$^{4}$
Jordan D. Collier,$^{5,6,7}$
Richard Dodson,$^{1}$
Simon P. Driver,$^{1}$\newauthor
Benne W. Holwerda,$^{8}$
Matt J. Jarvis,$^{3,9}$
Kenda Knowles,$^{10}$
Claudia Lagos,$^{1,2}$
Natasha Maddox,$^{11}$\newauthor
Martin Meyer,$^{1}$
Aaron S. G. Robotham,$^{1}$
Sambit Roychowdhury,$^{12}$
Kristof Rozgonyi,$^{12}$\newauthor
Nicholas Seymour,$^{13}$
Malgorzata Siudek,$^{14,15}$
Matthew Whiting,$^{7}$
Imogen Whittam,$^{3,9}$\\
$^{1}$ICRAR, The University of Western Australia, 35 Stirling Highway, Crawley, WA 6009, Australia \\
$^{2}$ARC Centre of Excellence for All Sky Astrophysics in 3 Dimensions (ASTRO 3D), Australia\\
$^{3}$Oxford Astrophysics, Denys Wilkinson Building, University of Oxford, Keble Rd, Oxford, OX1 3RH, UK\\
$^{4}$INAF -- Osservatorio Astronomico di Brera, via Brera 28, I-20121, Milano, Italy \& via Bianchi 46, I-23807, Merate, Italy\\
$^{5}$The Inter-University Institute for Data Intensive Astronomy (IDIA), Department of Astronomy, University of Cape Town,\\ Private Bag X3, Rondebosch, 7701, South Africa\\
$^{6}$School of Science, Western Sydney University, Locked Bag 1797, Penrith, NSW 2751, Australia\\
$^{7}$CSIRO, Space and Astronomy, PO Box 1130, Bentley, WA, 6102, Australia\\
$^{8}$University of Louisville, Department of Physics and Astronomy, 102 Natural Science Building, Louisville, KY 40292 USA\\
$^{9}$Department of Physics and Astronomy, University of the Western Cape, Robert Sobukwe Road, Bellville 7535, South Africa\\
$^{10}$Centre for Radio Astronomy Techniques and Technologies, Department of Physics and Electronics, Rhodes University, P.O. Box 94,\\ Makhanda 6140, South Africa\\
$^{11}$School of Physics, H.H. Wills Physics Laboratory, Tyndall Avenue, University of Bristol, Bristol, BS8 1TL, UK\\
$^{12}$University Observatory, Faculty of Physics, Ludwig-Maximilians-Universit\"{a}t, Scheinerstr. 1, 81679 M\"{u}nchen, Germany\\
$^{13}$International Centre for Radio Astronomy Research, Curtin University, Bentley, WA 6102, Australia\\
$^{14}$Institute of Space Sciences (ICE, CSIC), Campus UAB, Carrerde Can Magrans, s/n, 08193 Barcelona, Spain\\
$^{15}$Institut de Física d’Altes Energies (IFAE), The Barcelona Institute of Science and Technology, 08193 Bellaterra (Barcelona), Spain\\
}

\date{Accepted XXX. Received YYY; in original form ZZZ}

\pubyear{2023}

\begin{document}
\label{firstpage}
\pagerange{\pageref{firstpage}--\pageref{lastpage}}
\maketitle

\begin{abstract}
\label{sec:abstract}
The tight relationship between infrared luminosity (\LTIR) and 1.4\,GHz radio continuum luminosity (\Lradio) has proven useful for understanding star formation free from dust obscuration. Infrared emission in star-forming galaxies typically arises from recently formed, dust-enshrouded stars, whereas radio synchrotron emission is expected from subsequent supernovae. By leveraging the wealth of ancillary far-ultraviolet\,--\,far-infrared photometry from the Deep Extragalactic VIsible Legacy Survey (DEVILS) and Galaxy and Mass Assembly (GAMA) surveys, combined with 1.4\,GHz observations from the MeerKAT International GHz Tiered Extragalactic Exploration (MIGHTEE) survey and Deep Investigation of Neutral Gas Origins (DINGO) projects, we investigate the impact of timescale differences between far-ultraviolet\,--\,far-infrared and radio-derived star formation rate (SFR) tracers. We examine how the SED-derived star formation histories (SFH) of galaxies can be used to explain discrepancies in these SFR tracers, which are sensitive to different timescales. Galaxies exhibiting an increasing SFH have systematically higher \LTIR and SED-derived SFRs than predicted from their 1.4\,GHz radio luminosity. This indicates that insufficient time has passed for subsequent supernovae-driven radio emission to accumulate. We show that backtracking the SFR(t) of galaxies along their SED-derived SFHs to a time several hundred megayears prior to their observed epoch will both linearise the SFR-\Lradio relation and reduce the overall scatter. The minimum scatter in the SFR(t)-\Lradio is reached at 200\,--\,300\,Myr prior, consistent with theoretical predictions for the timescales required to disperse the cosmic ray electrons responsible for the synchrotron emission.
\end{abstract}





\clearpage
\section{Introduction}
\label{sec:intro}
Observing the radiation emitted across specific regions of the electromagnetic spectrum and studying their interrelationships is the cornerstone for understanding the physical processes taking place within galaxies. One such important relationship links the observed infrared (IR) and radio continuum emission (commonly observed at 1.4\,GHz), more generally referred to as the \textit{infrared-radio correlation} \citep*[IRRC; ][]{vanderKruit1971,vanderKruit1973,DeJong1985}. This correlation has been shown to hold over more than three orders of magnitude \citep{Helou1985,Condon1992,Yun2001}, displaying a confoundingly tight scatter of 0.2\,--\,0.3\,dex \citep{Sargent2010, Murphy2011, Delhaize2017, Molnar2021, Delvecchio2021}. The IRRC is believed to arise from the fact that both IR and radio continuum from galaxies in the absence of an active galactic nucleus (AGN) arises from a common origin --- namely star formation \citep{Condon1992, Kennicutt1998, Charlot&Fall2000}. Infrared emission ($\lambda=$\,8\,--\,1000\,\microns) comes predominantly from massive ($\gtrsim 5$\,\Msun) type O or early-type B stars whose intense ultraviolet (UV) radiation heats their surrounding dust, subsequently re-radiating at much lower frequencies in the infrared \citep{Sauvage2005, DaCunha2008, Xilouris2012, Bianchi2022}. Conversely, the radio emission is mostly dominated (at 1.4\,GHz) by the supernova explosions of more massive stars ($\gtrsim 8$\,\Msun) accelerating cosmic ray electrons (CRe) up to relativistic speeds, emitting synchrotron radiation as they follow galactic magnetic fields \citep{Condon1992, Bell2003, Murphy2009, Murphy2011}. Note, however, that whilst non-thermal mechanisms dominates radio emission from $\sim$1\,--\,30\,GHz, thermal free-free emission contributes approximately $\sim$10\,\% at 1.4\,GHz \citep{Condon&Yin1990, Condon1992, Rabidoux2014}.

The IRRC presents itself in a range of galaxy types, including late-type galaxies (\citealt*{Dickey&Salpeter1984}, \citealt{Helou1985}), dusty star-forming galaxies \citep{Giulietti2022}, compact dwarf galaxies \citep{Hopkins2002}, early-type galaxies \citep{Wrobel1988, Wrobel1991, Brown2011} with modest levels of star formation and, in some cases, merging systems \citep{Domingue2005}. Whilst historically the radio emission has been measured at 1.4\,GHz, the relation persists at lower radio frequencies, evidenced by observations with the GMRT\footnote{Giant Metrewave Radio Telescope} at 610\,MHz \citep{Ocran2020} and with LOFAR \footnote{Low Frequency Array} at 150\,MHz \citep{Smith2014, CalistroRivera2017, Gurkan2018, Wang2019, Bonato2021, Smith2021}. There is evidence that this correlation remains out to redshifts as far as $z\sim$\,3.5\,--\,4 \citep{Ibar2008, Delvecchio2021}, including studies of strongly gravitationally lensed systems \citep{Stacey2018, Giulietti2022}. 

Typically, the IRRC is parameterised by the logarithm of the ratio between the infrared and radio luminosities, \qIR, as per the following equation:

\begin{equation}
    q_\mathrm{IR} = \log\left(\frac{L_\mathrm{IR}\mathrm{[W]}}{3.75\times10^{12}\,[\mathrm{Hz}]}\right) - \log\left(L_\mathrm{1.4\,GHz}\,\mathrm{[W\,Hz^{-1}]}\right).
    \label{eqn:qTIR}
\end{equation}

Seminal work from \citet{Bell2003} using the total infrared luminosity (TIR; $\lambda=$\,8\,--\,1000\,\microns) observations found a constant value of \qTIR$ = 2.64$, whereas more recent studies have shown that this value may vary as a function of various galaxy properties. For example, a slight but statistically significant decrease of \qIR is seen with increasing redshift \citep{Seymour2009, Ivison2010, Basu2015}, often of the form of \qFIR$ \propto (1+z)^{[-0.2\,:\,-0.1]}$ (see e.g., \citealt{Magnelli2015, Delhaize2017, Ocran2020}). Several other studies using different datasets have not come to the same conclusion, noting a lack of any significant redshift-dependent evolution over $0 < z < 2$ \citep{Garrett2002, Ibar2008, Jarvis2010, Sargent2010, Smith2014, Bonato2021}. Other studies have found that \qTIR instead evolves predominantly with stellar mass \citep{Gurkan2018, Delvecchio2021, Molnar2021}, making the case that any subsequent evolution observed in redshift is likely the result of a biased sample selection whereby low-mass galaxies become increasingly under-represented with increasing redshift in flux-limited samples. \\

The linearity in this relation is perhaps surprising given that it assumes galaxies to be completely optically thick to UV emission \citep{Holwerda2005, Keel2014}. However, particularly at low stellar masses, galaxies are significantly less metal-rich \citep{Tremonti2004}, suggesting that this assumption may not hold. This low dust attenuation would result in underestimated infrared luminosities with respect to a given star formation rate. Coincidentally, cosmic rays accelerated during supernovae are assumed to lose all of their energy through synchrotron emission well before they have escaped the galaxy --- an assumption that may also break down in low-mass galaxies due to their lower gravitational potential \citep{Bourne2011}. The nature of the IRRC is thus still strongly debated, driven in part by a need to explain its surprisingly tight scatter and apparent linearity over several orders of magnitude. Some explanations require modelling the IRRC according to a \emph{non-calorimetric} \citep{Voelk1989} and optically-thin scenario \citep{Helou1993, Bell2003, Basu2015}, arguing that underestimates in both SFRs probed by infrared and radio emission in low stellar galaxies lead to a conspiracy in the linearity of the resulting IRRC \citep{Lacki2010a}.

Larger samples and more accurate photometry have since allowed the IRRC to be characterised in greater detail, showing that its power-law slope is slightly sub-unity \citep{Hodge2008, LoFaro2015, Basu2015, Davies2017, Molnar2021, Delvecchio2021} and may present a non-linearity at faint luminosities \citep{Hodge2008, Gurkan2018}. Understanding why is key to uncovering the origin of this correlation and indeed crucial for using observed radio luminosities as a proxy for star formation. Several models have been proposed to explain the non-linearity of the IRRC with early explanations suggesting a two-component model for infrared emission, including a warm ``active'' component borne out of the massive young stars heating dusty regions and a cooler component heated by an interstellar radiation field providing relatively constant cirrus emission \citep{Helou1986, LonsdalePersson1987, Fitt1988}. 

Furthermore, non-thermal emission from CRes may not account for their total energy as some fraction may have been able to escape before radiating all of their energy as synchrotron emission \citep{Niklas1997, Bell2003, Lacki2010a, Basu2015}. Young CRes driven away from star-forming regions may also lose energy via many different cooling mechanisms, including inverse Compton scattering, Bremsstrahlung, ionisation and adiabatic expansion. Some works have also attempted to account for the diminishing infrared emission found in low-mass star-forming galaxies (SFGs) with a complement of UV photometry \citep{Bell2005, Papovich2007, Barro2011, Davies2016, Delvecchio2021}, giving a holistic view of the SFR over short-to-moderate timescales (10\,--\,100\,Myr). This act of balancing the dust-obscured UV emission with the subsequently re-emitted infrared emission from the enshrouded dust effectively forms the basis of fitting the spectral energy distributions (SED) in galaxies \citep{DaCunha2008, Boquien2019, Robotham2020}. Crucially, however, radio continuum emission exhibits negligible attenuation from dust at frequencies of $\nu < 30$\,GHz, making it a useful star formation rate indicator in dusty SFGs \citep{Bell2003, Lacki2010b, Murphy2011, Kennicutt&Evans2012, Davies2017, Leslie2020}, particularly at high redshifts where dust attenuation is highly uncertain.\\

As a natural consequence of having such a tight relation for SFGs, studies have also identified populations of radio-bright AGN and radio-quiet quasars via an excess in their 1.4\,GHz radio emission with respect to their observed optical through infrared star formation rates \citep{Donley2005, Norris2006, Park2008, DelMoro2013, Bonzini2015, White2015, White2017, Thorne2022b}. However, up until recently, most radio continuum surveys have been hindered by shallow depths or narrow sky coverage, requiring stacking over populations of galaxies to reach a sufficiently high signal-to-noise to study the faint synchrotron emission of `normal' star-forming galaxies \citep{Davies2017, Delvecchio2021}. Upcoming and currently ongoing radio continuum surveys such as those underway with the Australian Square Kilometre Array Pathfinder \citep[ASKAP; ][]{Johnston2007, Johnston2008, Hotan2021}, Meer Karoo Array Telescope \citep[MeerKAT; ][]{Jonas2016} and Low-Frequency Array \citep[LOFAR; ][]{vanHaarlem2013} are now beginning to reveal the previously undetectable population of faint radio sources dominated by star formation processes \citep{Norris2011, vanderVlugt2021, Tasse2021, Sabater2021, Heywood2022}. This has opened an additional and complimentary avenue for measuring how star formation evolves in galaxies over cosmic time.\\

One aspect of the IRRC that has not been thoroughly explored is the fact that UV\,--\,IR emission and radio continuum emission likely probe SFR on different timescales. For example, \citet{Magnelli2015} found a moderate increase of 0.2\,dex in \qFIR for ``star-bursting'' galaxies above the main sequence of star formation. An empirical correction for this increase in \qTIR had previously been hinted at through a simple stellar evolution model by \citet{Biermann1976} that depended upon their $B$-band/radio ratio. Galaxies needing the largest corrections generally have the lowest radio luminosities but show bluer optical colours, suggesting that their current star formation rates may be lower than the average over the last $\sim$1 Gyr \citep{Condon1992}. These findings suggest that an aspect of the IRRC that warrants further investigation is the fact that the infrared and radio emission borne out of star formation probe different timescales.

More recently, \citet{Arango-Toro2023} showed that star formation rates derived from 1.4\,GHz emission could be significantly over-estimated due to galaxies with declining SFHs. The authors showed that this discrepancy can be resolved by backtracking the SFR of galaxies to several 100 Myrs prior. This can be understood by considering that the UV\,--\,IR emission that informs measurements of SFR typically occurs on shorter timescales than synchrotron from subsequent supernovae. Whilst infrared emission from dusty star-forming regions can emerge over mid-to-long timescales and can be slow to dissipate once star formation is suppressed \citep{Kennicutt1998}, ultraviolet emission traces much shorter timescales of $\sim$10\,Myr \citep{Grootes2017}. On the other hand, radio synchrotron emission will emerge once the CRes have propagated throughout a galaxy and are accelerated to sufficient relativistic velocities via Fermi acceleration. A seminal review of the radio emission in star-forming galaxies by \citet{Condon1992} showed that at an emitted frequency of $\nu_\mathrm{c}=1.4$\,GHz and for a magnetic field strength of $B=5\,\mathrm{\mu G}$, the typical lifetimes of synchrotron emitting electrons with isotropically distributed velocities is of order 100\,Myrs. Thus, if the star formation rates of galaxies vary significantly over these timescales (e.g. due to a starburst), the emission measured in the UV\,--\,IR and radio regimes will probe different epochs of star formation.

In this work, we attempt to quantify the impact of mismatched timescales between infrared and radio continuum emission by combining the multi-wavelength surveys of both DEVILS and GAMA with the corresponding radio continuum observations from the MIGHTEE and DINGO surveys. The wealth of ancillary data in these surveys opens a previously untapped avenue for exploring how the radio properties of ``normal'' star-forming galaxies are impacted by a multitude of galaxy properties (stellar masses, star formation rates, metallicities, AGN fractions, etc.) as derived from modelling their spectral energy distributions. Understanding these dependencies will become essential as radio continuum observations become ubiquitous and used as a routine SFR indicator \citep[see e.g.,][]{Murphy2009, Schober2022} in the upcoming era of the Square Kilometre Array Observatory \citep[SKAO;][]{Dewdney2009} and next-generation Very Large Array \citep[ngVLA; ][]{Murphy2018} telescopes. A key scientific goal of these telescopes is to measure the cosmic star formation history using the radio continuum as a dust-unbiased tracer of star formation (\citealt*{Ciliegi&Bardelli2015}; \citealt{Jarvis2015}). \\

This paper is structured as follows. A brief description is given in  Section\,\ref{sec:data_and_sample} for each of the multi-wavelength surveys of DEVILS and GAMA, as well as the overlapping radio continuum observations from MIGHTEE and DINGO followed by a discussion on the sample selection and data being used. In Section\,\ref{sec:results}, we explore the infrared-radio correlation and radio star formation rate calibrations as presented by these surveys and compare them with previous studies. We then explore the impact of the mismatch in timescales between star formation rates as derived from UV\,--\,FIR wavelengths and the radio continuum in Section\,\ref{sec:analysis}. We discuss the implications with respect to star formation histories in Section\,\ref{sec:discussion} and summarise our findings in Section\,\ref{sec:conclusions}. Throughout this paper, we assume a \citet{Chabrier2003} initial mass function (IMF) and magnitudes are given based on the AB system, also adopting cosmological parameters from the \citet{PlanckCollaboration2016}, namely $H_{0} = 67.8\,\mathrm{km\,s^{-1}\,Mpc^{-1}}$, $\Omega_\mathrm{M} = 0.308$ and $\Omega_\mathrm{\Lambda} = 0.692$.

\section{Observational Data and Sample Selection}
\label{sec:data_and_sample}

\begin{figure*}
    \centering
    \includegraphics[width=0.32\textwidth]{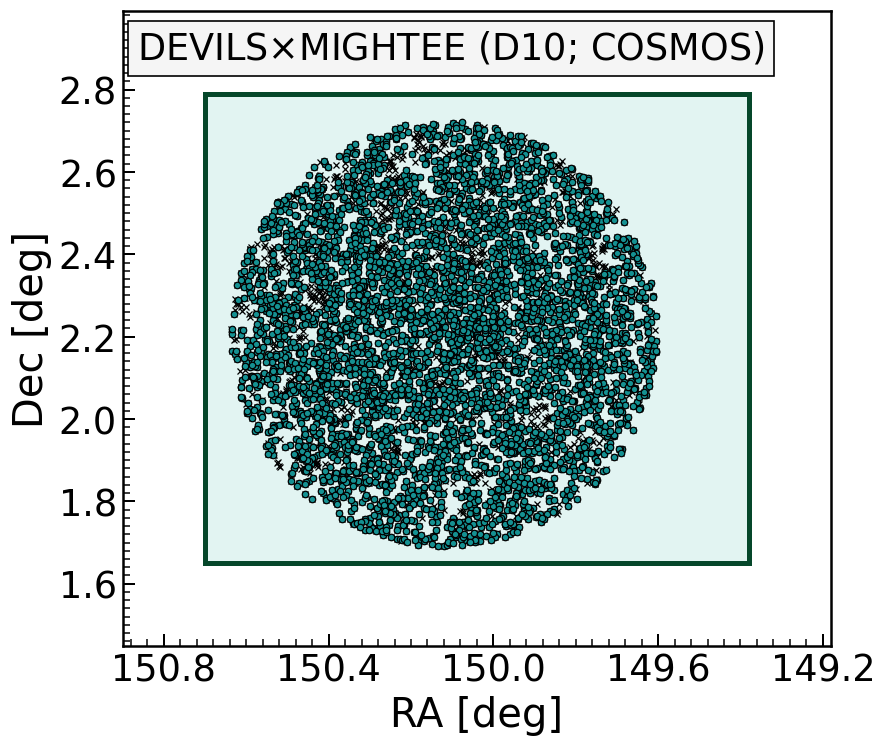}
    \includegraphics[width=0.52\textwidth]{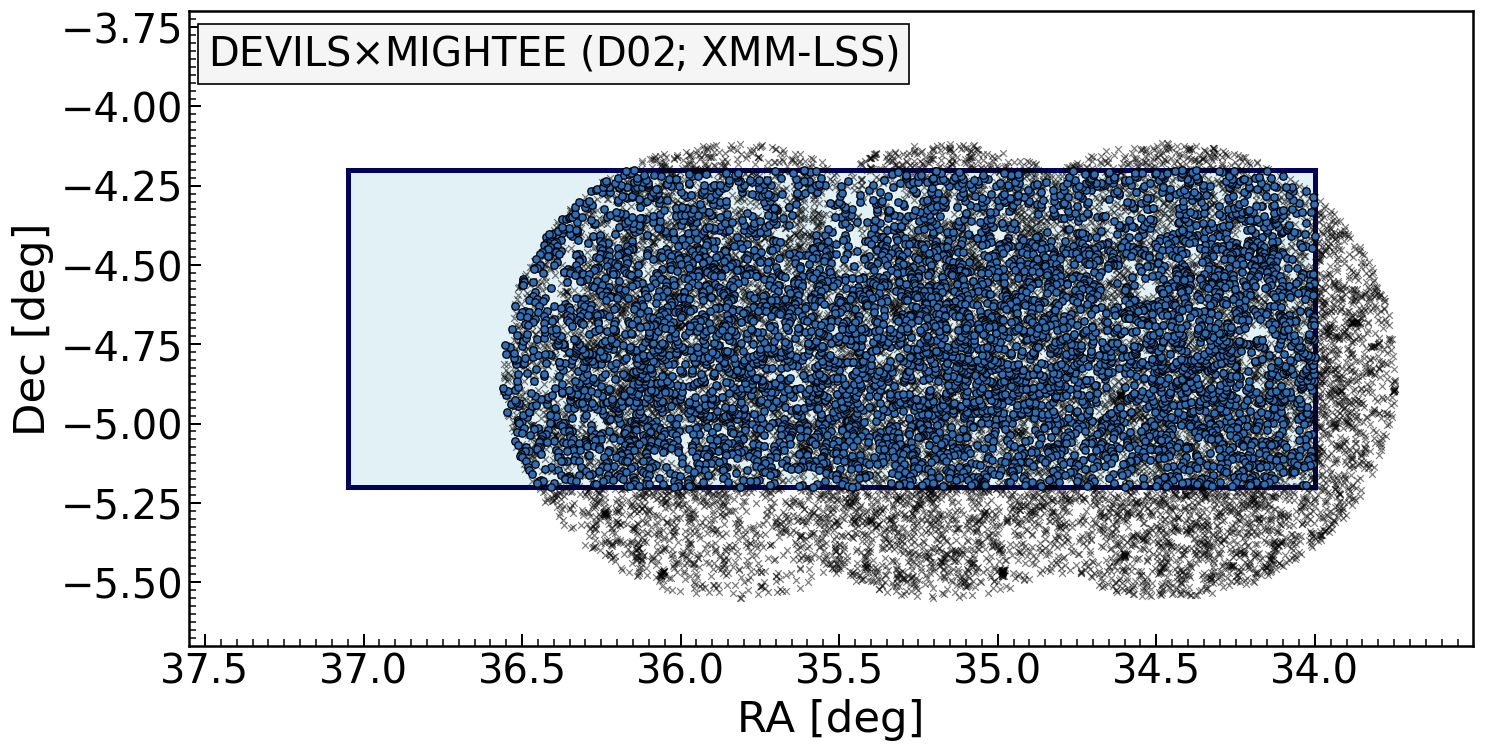}
    \includegraphics[width=0.49\textwidth]{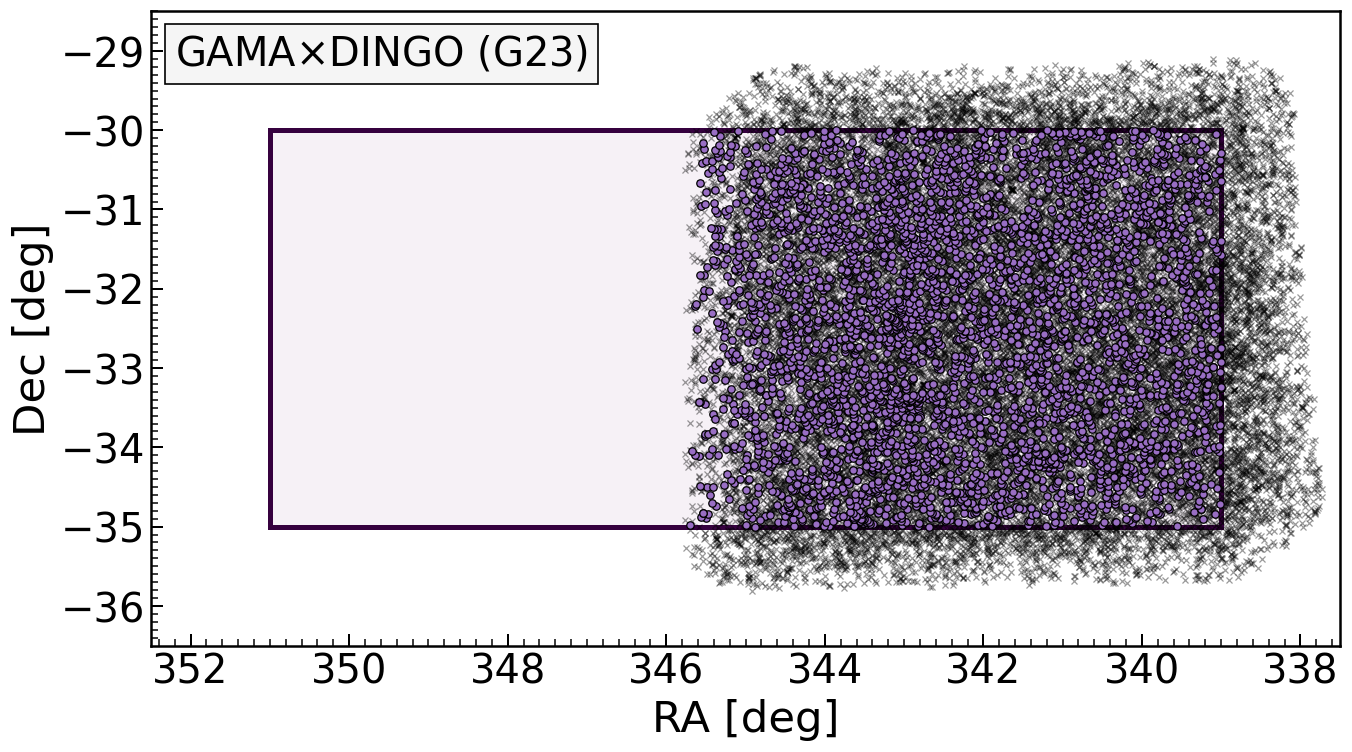}
    \includegraphics[width=0.49\textwidth]{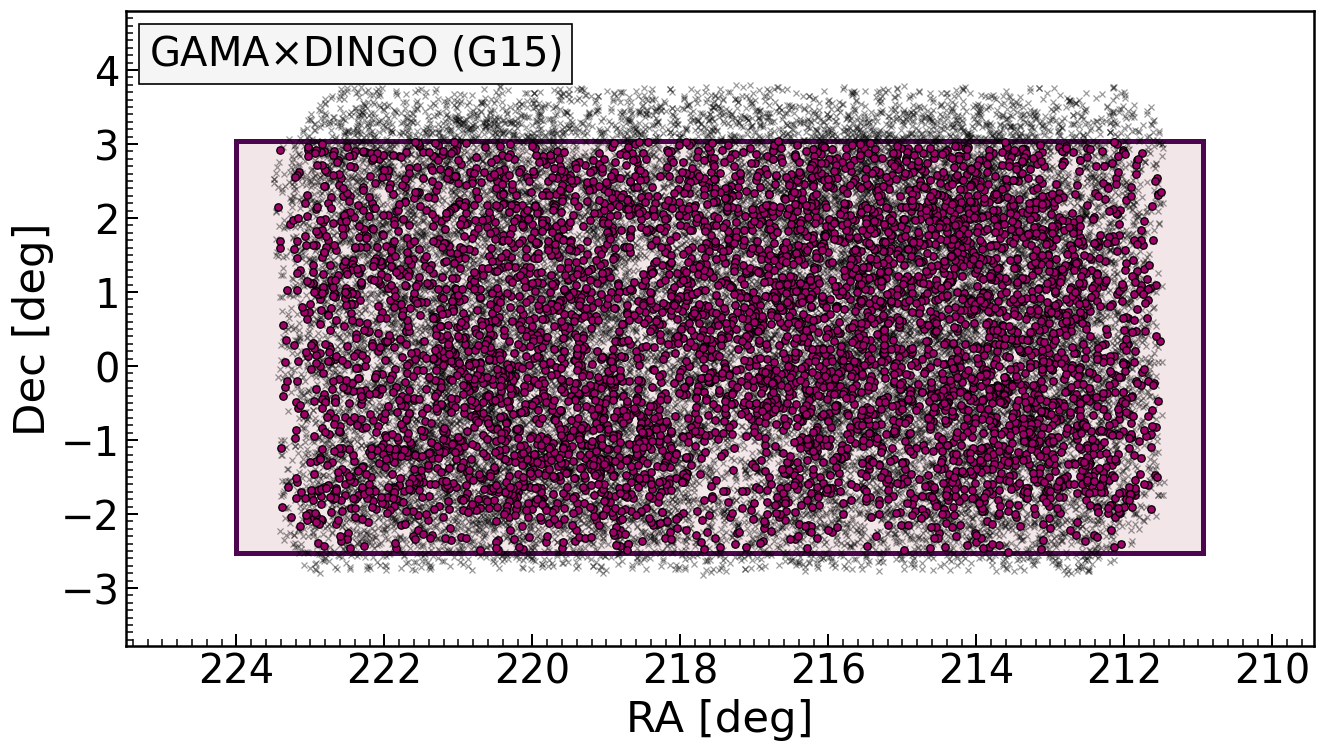}
    \includegraphics[width=0.49\textwidth]{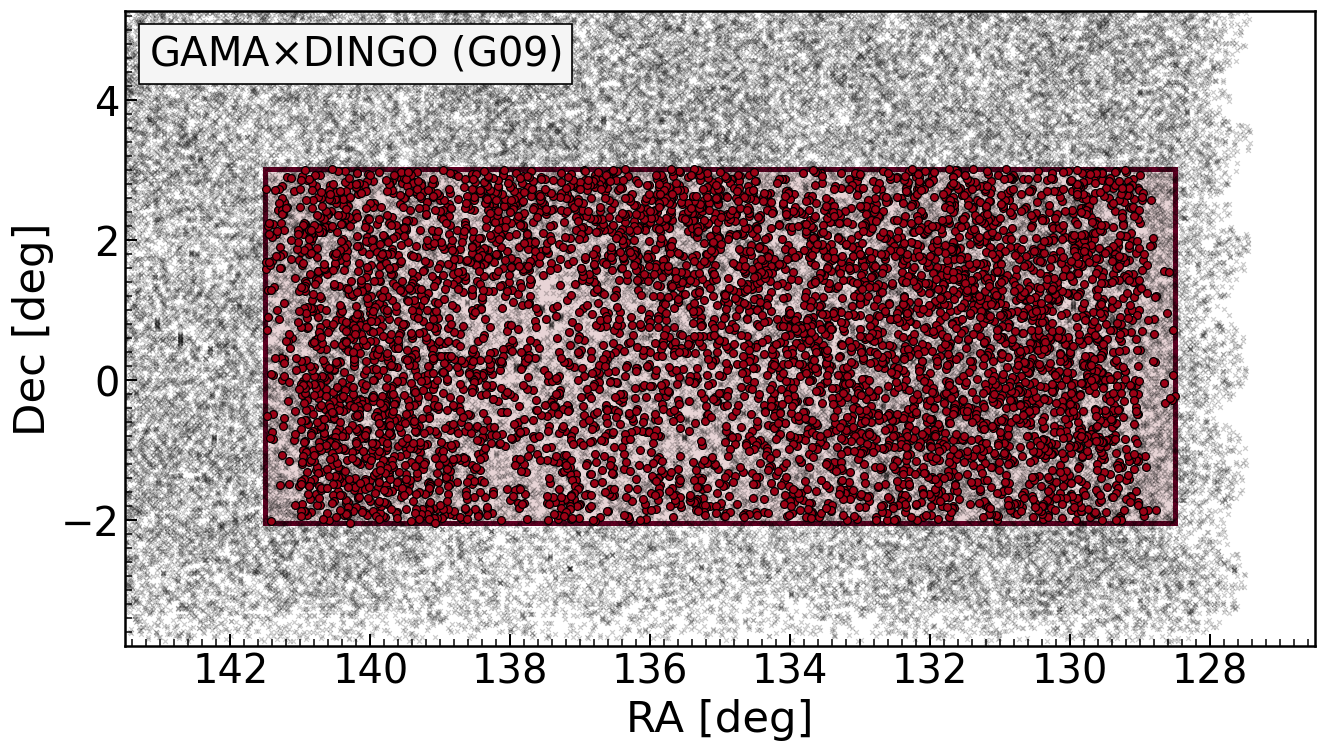}
    \caption{The overlap regions of the data sets used in this work, including the DEVILS$\times$MIGHTEE cross-matched regions of the COSMOS (D10) and XMM-LSS (D02) in the top panels as well as the GAMA$\times$DINGO cross-matched regions of the G23, G15 and G09 fields. In each panel, the outlined regions show the coverage of the DEVILS and GAMA multi-wavelength datasets and the coloured circles show the subsets of those sources that have 1.4\,GHz radio detections in their corresponding radio continuum surveys. Grey crosses show radio detections without a corresponding detection at optical wavelengths.}
    \label{fig:fields_sky_coverage}
\end{figure*}

\subsection{Multi-wavelength Data from the Far-UV to Far-IR}
\label{sec:mw_data}
In this section, we summarise the multi-wavelength photometric and spectroscopic catalogues that form the basis of the galaxy samples used in this study.

\subsubsection{DEVILS}
\label{sec:mw_data.DEVILS}

The Deep Extragalactic Visible Legacy Survey \citep[DEVILS; ][]{Davies2018} is a spectroscopic campaign conducted, in part, on the Anglo-Australian Telescope (AAT) with the goal to increase the redshift completeness of the currently undersampled epochs at intermediate redshifts (0.3\,$< z <$\,1.0). The design at the inception of the survey \citep{Davies2018} was to compile a large, spectroscopically complete sample down to a limiting brightness of $Y$\,$<$\,21\,mag in three deep extragalactic fields: COSMOS (D10), XMM-LSS (D02) and ECDFS\footnote{The Cosmic Evolution Survey field, The XMM Large Scale Structure and Extended Chandra Deep Field South, respectively.} (D03). These fields were chosen specifically for their wealth of deep, panchromatic imaging data sets collected from various ground- and space-based telescope facilities. The crucial characteristic of DEVILS is that it has been supplemented by pre-existing catalogues of spectroscopic and photometric redshift measurements, complemented by the spectroscopic observations made using the AAOmega fibre-fed spectrograph \citep{Saunders2004, Sharp2006} on the AAT. The result of these efforts is a deep, highly spectroscopically complete ($>85$\,\%) sample of $\sim$50,000 galaxies down to a $Y$-band magnitude of $<$\,21.0\,mag in 3\,\degsq collectively over a redshift range of 0.0\,$< z <$\,1.0. The automatic source-finding and image analysis package \ProFound \citep{Robotham2018} was paramount in standardising the extraction of photometry across all bands from the FUV to FIR. It has been shown that applying these consistent data processing and analysis methods to this comprehensive collection of up-to-date datasets is crucial to minimising the non-negligible and often overlooked errors that arise from inconsistencies in selection methods, magnitude zero-point offsets and photometric measurement techniques (see \citealt{Davies2021} for details).

Most of the current scientific application of DEVILS data has come from the D10 field due, in part\footnote{Additionally, shutdowns and unexpected performance losses at the AAT have also refocused DEVILS observations to the completion of D10 first.}, to the superior multi-wavelength photometry (specifically UltraVISTA and HSC observations) and redshift measurements available in this field. However, in order to maximise our sample size, we use both the D10 and D02 fields (see Figure\,\ref{fig:fields_sky_coverage}) as both the multi-wavelength photometry and 1.4\,GHz radio continuum observations are available in these fields (see Table \ref{tab:data_summary} for a summary of the basic observing properties in these fields). The D10 and D02 fields contain 493,627 and 302,615 galaxies, respectively, with sufficient multi-wavelength coverage to fit an SED model and were not labelled as stars (\texttt{starflag}), artefacts (\texttt{artefactflag}) or masked (\texttt{mask}) according to \citet{Davies2021}. As discussed in \citet{Thorne2021}, the redshifts for the DEVILS catalogue are taken as the best available from a range of spectroscopic, grism and photometric sources --- the typical photometric redshift accuracy is $\lesssim0.01$ for the majority of the sample.

\subsubsection{GAMA}
\label{sec:mw_data.GAMA}
The Galaxy And Mass Assembly \citep[GAMA; ][]{Driver2011, Hopkins2013, Liske2015, Driver2022} survey is a legacy campaign to obtain redshifts using the AAT and covering a total of $\sim$\,250\,\degsq over its three equatorial regions: G09, G12 and G15\footnote{As with DEVILS notation, the integer represents the approximate right ascension of that field in hours.} as well as the G02 field located at a declination of $\delta\,\sim\,-7^\mathrm{\circ}$ and G23 field located at $-32.5^\mathrm{\circ}$ in the Southern Galactic Cap. The latter is the foremost GAMA field and will be the target of various southern hemisphere surveys including the forthcoming Wide Area VISTA Extragalactic Survey \citep[WAVES;][]{Driver2019} and --- crucially for the work presented here --- recent observations from the Deep Investigations of Neutral Gas Origins \citep[DINGO; ][]{Meyer2009, Rhee2023}, discussed in detail in Section \ref{sec:radio_data.DINGO} below. The G23 field has a spectroscopic survey limit of $i < 19.2$\,mag \citep{Liske2015, Driver2022} at a spectroscopic completeness of 90\,\%. To date, few studies have made use of the G23 field, with notable exceptions of \citet{Bilicki2018, Vakili2019}, in part due to the recent assimilation of deeper homogeneous $ugri$ imaging from the ESO VST Kilo-Degree Survey \citep[KiDS;][]{Kuijken2019}. However, the G23 field is becoming popular for ASKAP early science and pilot surveys \citep[e.g.][]{Leahy2019, Allison2020, Gurkan2022, Rhee2023}. With the inclusion of the KiDS data, \citet{Bellstedt2020a} re-derived the optical/near-IR catalogues within the GAMA regions using a uniform approach based on the source-finding tool \ProFound \citep{Robotham2018}. This involved collating images from \textit{GALEX} \citep{Zamojski2007}, VST KiDS \citep{deJong2013}, VISTA VIKING \citep{Arnaboldi2007}, \textit{WISE} \citep{Wright2010}, and \textit{Herschel} \citep{Pilbratt2010} imaging campaigns.

In total, the GAMA survey contains redshifts for 330,000 galaxies across its five sky regions. In this work, we use the three fields of G23, G15 and G09 (see Figure\,\ref{fig:fields_sky_coverage}), which for a 95\,\% spectroscopic completeness yield 45,427, 73,842 and 68,959 sources, respectively, prior to cross-matching with DINGO continuum sources. Table\,\ref{tab:data_summary} summarises the number of galaxies in each field that have been modelled using \ProSpect including AGN templates as described in \citet{Thorne2022b}. The spectroscopic and photometric measurements that underpin the derivation of physical galaxy properties have been extracted in a consistent manner in each of the DEVILS and GAMA fields. This is crucial for this and future works capitalising on comparisons between the often disparate redshift regimes covered by local surveys with the distant Universe, whereby unforeseen biases can be introduced if not treated in a consistent manner.

\subsection{Radio continuum data}
\label{sec:radio_data}
Below, we briefly outline the radio continuum data from the two SKAO precursor instruments, MeerKAT and ASKAP. The relevant surveys conducted on these instruments are MIGHTEE and DINGO, which respectively complement the multi-wavelength surveys of DEVILS and GAMA.

\subsubsection{MIGHTEE}
\label{sec:radio_data.MIGHTEE}
The MeerKAT International Gigahertz Tiered Extragalactic Exploration \citep[MIGHTEE; ][]{Jarvis2016, Heywood2022} survey is one of the MeerKAT telescope's largest ongoing surveys, collecting $\sim$1000 hours of radio continuum, polarimetry (Sekhar et al. in prep.) and 21-cm emission line \citep{Maddox2021} observations in the L-band (856\,--\,1712\,MHz) and S-band (2\,--\,4\,GHz). Four extragalactic deep fields make up the collective 20\,\degsq of observations including three fields that coincide with DEVILS, including COSMOS, XMM-LSS and ECDFS as well as an additional field from the southernmost region of the European Large Area ISO Survey Southern field (ELAIS-S1). In this work, we focus only on the first two fields, where observations are completed and early science products available. The top two panels of Figure \ref{fig:fields_sky_coverage} show the overlap regions between the DEVILS and MIGHTEE early science data in the D10--COSMOS and D02--XMM-LSS fields, respectively.

Full details of the MIGHTEE Early Science data can be found in \citet{Heywood2022}, but Table \ref{tab:data_summary} summarises the relevant properties of the radio continuum observations used in this work. MIGHTEE Early Science data offers the two weighting schemes for robust values of 0.0 and -1.2 as a trade-off between sensitivity and angular resolution. We employ the robust 0.0 weighted images, which achieve an RMS thermal noise of 1.7\,\muJybeam with a resolution of 8.6$''$ and 8.2$''$ in the COSMOS and XMM-LSS fields\footnote{In fact, due to the three overlapping pointings in the XMM-LSS field, the deepest combined areas reach thermal noise levels of 1.5\,\muJybeam, with a measured rms level noise of 6.0\,\muJybeam.}, respectively. Note, however, that the robust 0.0 images are fundamentally limited not by thermal noise but by the classical confusion limit at 4.5\,\muJybeam, which equates to the surface density where point sources can no longer be reliably separated \citep{Heywood2013}. This weighting scheme was selected primarily because we are interested in detecting generally-fainter radio continuum emission from star-forming galaxies rather than bright active galactic nuclei and are generally agnostic to the underlying small-scale structures present in these sources.

Note that because of the frequency dependence of the primary beam as well as the wide bandwidth, the effective frequency gradually decreases outwards from the centres of the pointings \citep{Heywood2022}. This is resolved by rescaling the flux density of detected sources to a common effective frequency of 1.4\,GHz assuming a spectral index\footnote{$S_\mathrm{\nu}\,\propto\,\nu^{-\alpha}$; where $S_\mathrm{\nu}$ is the integrated flux density at a given frequency $\nu$.} of $\alpha = 0.7$, which is commensurate with several studies using L-band observations \citep{Smolcic2017a, CalistroRivera2017, An2021, Hale2023}.

\subsubsection{DINGO}
\label{sec:radio_data.DINGO}
The Deep Investigations of Neutral Gas Origins (DINGO) is a deep 21\,cm spectral line survey using ASKAP. Note that for this work, we do not use the spectral line cubes, instead utilising the measurements of the underlying radio continuum. This further highlights the power of the next generation of sensitive radio telescope arrays such as ASKAP and MeerKAT to both obtain large numbers of radio source counts out to the high redshift Universe as well as commensurately measuring well-resolved spectral line emission from neutral hydrogen (\HI) down to low column densities of $\lesssim10^{20}$\,cm$^{-2}$.

The DINGO pilot survey observed the GAMA fields of G15 and G23 with the full array of 36 ASKAP antennas and the full 288\,MHz bandwidth (15,552 channels at 18.5\,kHz channel resolution). These were completed in the ASKAP receiver band 2 with observing frequency ranges of 1.146\,--\,1.434\,GHz for G15 and 1.152\,--\,1.440\,GHz for G23. Observations in these fields are separated into two adjacent tiles with 30\,\degsq fields of view. These tiles are comprised of two interleaving ASKAP pointings with beam footprints of 6$\times$6\,deg$^{2}$. The GAMA G09 field was observed separately as a follow-up to eROSITA observations by an ASKAP observatory project called Survey With ASKAP of GAMA-09 X-ray (SWAG-X). The observational parameters for this field were identical to G23 apart from the beam-forming configuration, instead using a close-packed 36 footprint of three interleaving pointings. A total of six tiles make up the SWAG-X observations, extending far beyond the nominal footprint of the G09 region. The G23 field has also been observed with sources catalogued as part of the ASKAP Evolutionary Map of the Universe (EMU) survey in \citet{Gurkan2022}, however, to avoid further systematic differences between fields, we simply use the DINGO observations in G23.

In these pilot survey data, the total integration time is $\sim$35.5\,hours ($\sim$17.7\,hours per tile) for G15, $\sim$47\,hours for the single tile in G23 and 104\,hours ($\sim$17.3\,hours per tile) for G09. This equates to an RMS noise level of 38 and 39\,\muJybeam in the G15 and G09 fields, whereas the deeper G23 observations get down to 18\,\muJybeam. The resolution varies between each field, with restoring beams of $7''\times6.6''$ in G23, $12.5''\times7.7''$ in G15 and $11''\times9.5''$ in G09.

A comprehensive description of the DINGO processing is given in \citet{Rhee2023}, but we provide a brief summary here focusing on the continuum data products, which are used in this work. Each data set yields a 72-beam (in G15 and G23) or 108-beam (in G09) combined continuum image. The observed data with 18.52\,kHz resolution are then averaged into a 1 MHz-wide channel for continuum imaging and self-calibration to correct for time-dependent phase errors. These images are then combined with the combined data products from the other tiles to cover the full field of view of each GAMA field. Then continuum source finding processing is applied to the entire continuum images using the \texttt{ASKAPsoft} \citep{Guzman2019} source finding task, called \texttt{selavy} \citep{Whiting2012}. 

Note, for DINGO observations, faint ($\lesssim$1\,mJy) unresolved sources detected by \ProFound often contain too few pixels to completely encompass the extent of the beam, thereby underestimating the true flux density of a source. To correct this, we replicate the correction presented in \citep{Hale2019}. For each source, we overlay its segmentation region over the reconstructed Gaussian beam and calculate the fraction of the total beam flux that is captured by the identified pixels. For bright and extended sources, the beam is completely covered by the segment, thus no correction is required. However, the correction can be as large as 0.5\,dex in the faintest sources detected, where very few pixels cover the beam. As the DINGO observation in G23 was exposed for $\sim3$ times longer than G15 and G09, a greater number of sufficiently faint sources are found that require a larger correction.

\setlength{\extrarowheight}{0.2cm}

\begin{table*}
    \centering
    \caption{An overview of the observing properties for both the multi-wavelength datasets (DEVILS and GAMA) and radio continuum surveys (MIGHTEE and DINGO) used throughout this work. For the multi-wavelength surveys, the columns show the sky coverage, limiting magnitude needed to achieve a 95\,\% spectroscopic completeness and the size of the initial sample, i.e., all sources with redshifts and SED modelled that were not masked or flagged for artefacts or stars. For the DEVILS fields, these numbers are separated further into the types of redshifts as either spectroscopic, grism or via photometric methods \citep[see][for the relevant sources of each]{Thorne2021}. For the radio continuum surveys, we show the noise properties from both the thermal noise and classical confusion as the latter is the more dominant source of noise in the lower resolution (Briggs robust 0.0 weighting) MIGHTEE images. The number of radio detections (SNR $> 3\,\sigma$) cross-matched within 3$''$ is given and in brackets are those that have been selected as both star-forming and show no sign of an AGN as described in Section\,\ref{sec:selecting_SFGs}. $^{\dagger}$ DEVILS observations of the XMM-LSS field achieved spectroscopic completeness in only the western half where the MIGHTEE pointings are also concentrated.}
    \begin{tabular}{p{2.5cm}x{2cm}x{4cm}x{3cm}x{1cm}x{1cm}x{1cm}}
        \hline
        \multirow{2}{*}{\shortstack{Multi-wavelength\\ surveys}} & \multirow{2}{*}{\shortstack{Sky coverage\\ (\degsq)}} & \multirow{2}{*}{\shortstack{Spectroscopic completeness\\ limit (mag)}} & \multirow{2}{*}{\shortstack{\# Initial galaxy\\ sample}} & \multicolumn{3}{c}{Redshift type} \\ \cline{5-7}
        & & & & photo & spec & grism \\ \hline
        \textbf{DEVILS} &  &  &  &  &  & \\
          \;\;D10 (COSMOS) & 1.5 & 20.5 ($Y$-band) & 493,627 & 462,240 & 24,086 & 7,301 \\
          \;\;D02 (XMM-LSS) & 3.0$^{\dagger}$ & 21.2 ($Y$-band) & 302,615 & 263,489 & 22,146 & 16,980 \\
        \textbf{GAMA} &  &  &  &  &  & \\
          \;\;G23 & 50 & 18.91 ($i$-band) & 45,427 &  &  & \\ 
          \;\;G15 & 50 & 19.16 ($i$-band) & 73,842 &  &  & \\
          \;\;G09 & 50 & 19.16 ($i$-band) & 68,959 &  &  & \\ \hline
        \multirow{2}{*}{\shortstack{Radio continuum\\surveys}} & \multirow{2}{*}{\shortstack{Sky coverage\\(\degsq)}} &  \multirow{2}{*}{\shortstack{1.4\,GHz thermal noise\\ $[$classical confusion limit$]$\\ (\muJybeam)}}  & \multirow{2}{*}{\shortstack{\# Cross-matched \\sample\\ $[$SFGs \& non-AGN$]$}} & \multicolumn{3}{c}{Redshift type} \\ \cline{5-7}
        &  &  &  & photo & spec & grism \\ \hline
        \textbf{MIGHTEE} &  &  &  &  &  & \\
          \;\;COSMOS & 1.5 & 1.7 $\left[\,4.5\,\right]$ & 975 [712] & 47 & 643 & 22 \\
          \;\;XMM-LSS & 3.5 & 1.5 $\left[\,4.5\,\right]$ & 2,766 [1,993] & 636 & 1,124 & 233 \\
        \textbf{DINGO} &  &  &  &  &  & \\
          \;\;G23 (tile 0 only) & 30 & 18 & 686 [597] &  &  & \\
          \;\;G15 & 60 & 38 & 982 [807] &  &  & \\
          \;\;G09 & 150 & 39 & 1,646 [1,421] &  &  & \\
        \hline
    \end{tabular}
    \label{tab:data_summary}
\end{table*}

\subsection{SED fitting of multi-wavelength data}
\label{sec:mw_data.sed_fitting}
Recently, SED fitting was performed using \ProSpect \citep{Robotham2020} to characterise the stellar population properties of galaxies in both the DEVILS \citep{Thorne2021, Thorne2022a} and GAMA \citep{Bellstedt2020b} regions. SED fits in XMM-LSS are forthcoming with the results being presented in this and future works. These multi-wavelength datasets typically consist of 22 bands in the DEVILS fields and 20 bands in the GAMA fields, extracted from a variety of facilities including \textit{GALEX}, CFHT \citep{Capak2007}, Subaru HSC \citep{Aihara2019}, VST, VISTA \citep{McCracken2012}, \textit{WISE}, \textit{Spitzer} \citep{Sanders2007, Laigle2016} and \textit{Herschel} (see \citealt{Thorne2022a}, \citealt{Bellstedt2020b} for details).

As illustrated in the above works and other studies \citep{Pacifici2023}, one of the key improvements made over previous SED fitting works is the addition of an evolving metallicity prescription and the implementation of star formation histories (SFHs) that can be flexibly defined by several functional forms or other non-parametric definitions. Briefly, the SFHs modelled in \citet{Bellstedt2021} and \citet{Thorne2021} use a skewed normal function with a truncation imposed such that $SFR\equiv0$ at the beginning of the Universe ($z=11.8$ as per these implementations). This has been shown to accurately model the variety of star formation histories observed across many classes of galaxies; see \citep{Robotham2020} for comparisons with simulated galaxies from the SHARK \citep{Lagos2018} semi-analytic model. For more details on the application of \ProSpect and implementations using other data sets, see \citet{Robotham2020} and \citet{Bellstedt2020b}. Here we use the \ProSpect implementation described in \citet{Thorne2022b}, which, in addition to the above prescriptions for star formation and metallicity histories, also incorporates an AGN template originally outlined by \citet{Fritz2006} and further expanded in \citet{Feltre2012}. Note, that in practice the SED-fitting is performed in an identical manner for both the GAMA and DEVILS datasets, allowing for a standardised comparison across a redshift baseline of $0 < z < 1$ (e.g., \citealt{DSilva2023}).

\subsection{Selecting star-forming galaxies}
\label{sec:selecting_SFGs}
Here we describe the process of selecting the subsets used in the analyses between the combined \DEVILSxMIGHTEE and \GAMAxDINGO data sets. Figure \ref{fig:fields_sky_coverage} shows the overlapping sky coverage shared between the multi-wavelength surveys of DEVILS and GAMA with the corresponding radio continuum surveys of MIGHTEE and DINGO, respectively. The early data release of MIGHTEE contains a single pointing completely encompassed by the 1.5\,deg$^{2}$ covered by the DEVILS D10 footprint, whereas the three MIGHTEE pointings in the XMM-LSS field only partially overlap ($\sim$2.6\,\degsq) with D02. Likewise, the early data release of DINGO bounds the western half of the G23 region.\\

In the COSMOS field, we make use of an early science MIGHTEE catalogue of cross-matches between radio sources and optical/near-IR detections to correctly assign radio sources to their optical counterparts \citep{Whittam2024}. Briefly, the cross-matching process involved visual inspection of MIGHTEE radio continuum contours over UVISTA K-band imaging to assign the most probable optical counterparts. In the remaining fields, we search for the nearest neighbour within a 3$''$ radius of the positions of sources in the multi-wavelength catalogues. Although we use the poorer resolution MIGHTEE maps, less than 5\% of sources are flagged as having potential source blending in \citet{Whittam2024}. Note that in addition to having poorer spatial resolution, the radio continuum emission can trace markedly different features and extents to the stellar light, such as the jets and lobes emanating from AGN. However, this radius is appropriate for SFGs, which are typically compact in radio continuum images at these redshifts. For the multi-wavelength data, we use the \texttt{RAmax} and \texttt{Decmax} parameters measured from the \ProFound source-finding catalogues.

We then apply stellar mass completeness cuts (see Section\,\ref{sec:sensitivities_in_samples}) and limit to redshifts $z < 1.0$ in DEVILS and $z < 0.5$ in GAMA as our samples are highly stellar-mass incomplete outside of these regimes. Table \ref{tab:data_summary} gives a summary of the number of detected galaxies in each survey as well as the total number of cross-matched pairs found. We also require a detection in at least one of the five Herschel FIR filters ($\lambda = $\,100\,--\,500\,\microns), ensuring reliable measurement of the SED-derived infrared luminosities and star formation histories, which are heavily constrained by the dust emission properties probed in the mid-to-far-infrared regime. The D02 field provides the greatest number of infrared- and radio-detected sources simply due to a larger sky coverage but generally contains poorer quality photometric redshifts than galaxies in the D10 field.

To understand the underlying scatter in the IRRC, it is important to select a sample of galaxies that are both actively star-forming and show no indication of AGN activity. Any contribution from AGN in the population of star-forming galaxies will result in an excess in the radio emission, hence be situated below the \LTIR-\Lradio relation. Indeed, it has become commonplace to associate a significant excess of radio emission from what would be expected from star-formation alone as a means of identifying AGN \citep{Donley2005, Norris2006, Park2008, DelMoro2013, Bonzini2015, Delvecchio2017}. To select SFGs, previous works have used cuts in $NUV-r$ colour \citep{Smolcic2017b, Delhaize2017, An2021} or optical versus infrared colours \citep{Davidzon2017, Delvecchio2017}, both of which tend to trace recent ($\lesssim 100$\,Myr) star formation. In this work, SFGs are selected above a limit of 0.5\,dex lower than the star-forming main sequence (SFMS) defined at the redshift of each source. The SFMS was parameterised in \citet{Thorne2021} according to a double power-law fit to the SFR-\Mstar plane in bins of roughly 700\,Myr in lookback time.

Finally, we use several distinct selection criteria across multiple wavelength regimes to minimise additional contamination from AGN, which are likely to increase \Lradio particularly towards higher stellar mass \citep{Best2012}, resulting in a less steep slope in the \LTIR-\Lradio relation. In each of the datasets, we make use of the \fAGN parameter derived from the AGN templates incorporated into the fitted SED model \citep{Thorne2022b}. The \fAGN quantity equates to the fraction of flux contributed by an AGN component between 5\,--\,20\,\microns, which encompasses the 6.2\microns polycyclic aromatic hydrocarbon (PAH) emission that is often associated with AGN activity \citep{Magdis2013, Dale2014}. For the purpose of this work, a value of \fAGN\,$> 0.1$ is taken to indicate a significant AGN component is present, following the definition of previous studies \citep{Leja2018}. \citet{Thorne2022b} showed that 91\,per cent of AGN selected via narrow and broad emission lines were found to have \fAGN$> 0.1$.

The above AGN classification potentially misses a population of low-excitation radio galaxies (LERGs), which do not always show a signature in their FUV\,--\,FIR SED; e.g. Hansen et al. (in prep.). Although such galaxies are unlikely to be star-forming (\citealt*{Heckman&Best2014}; but see also \citealt{Whittam2022}), we attempt to account for incorrectly classified non-AGN by taking the approach of \citet{CalistroRivera2017} to remove sources with an excess 1.4\,GHz radio luminosity with respect to their infrared luminosity. A cut is made to remove all sources that are $2\sigma$ lower than the mean \qTIR (see Table\,\ref{tab:LTIR_vs_L1.4GHz_params}). This \qTIR cut is used in addition to the primary \fAGN selection and while it is necessary to remove the few extreme radio-excess outliers, less conservative cuts do not significantly alter the slope of the relation. Ideally, one might avoid using the distribution of \qTIR as a means of removing AGN, however, we believe this secondary cut to be sufficiently conservative that it does not introduce a significant bias across the plane of the \LTIR-\Lradio relation. 

\citet{Whittam2024} capitalised on the wealth of multi-wavelength data within COSMOS to classify radio sources into SFGs and AGN across several different metrics based on distinct emission mechanisms expected if an AGN is present. The authors find that for sources where an optical counterpart is identifiable (i.e. in \citealt{Whittam2024}), 35\,per cent host an AGN. We implement their AGN selection criteria in the D10 sample, which includes various cuts based on the observed X-ray, optical, mid-IR and broad/narrow emission lines. In the region of overlap between these catalogues, we find that 80.0\,per cent of the SFG sample classified as non-AGN in \citet{Whittam2022} are also classified as non-AGN via the \fAGN$<0.1$ cut, which is used more generally in all other fields. On the other hand, 34\,\% of the galaxies flagged as AGN in \citet{Whittam2022} have \fAGN < 0.1, potentially leading to contamination of a star-forming sample if used in isolation of the aforementioned SFG and radio-excess cuts. See Appendix \ref{appndx:qTIR_distribution} for the distribution of \qTIR values for SFGs and AGN based on the above selection criteria. 

Table\,\ref{tab:data_summary} shows the initial number of galaxies in each of the fields separately, which when cross-matched with their corresponding radio continuum surveys, yield 31,548 sources with 1.4\,GHz detections. Of these, a further 16,925 are identified as star-forming galaxies with no indication of a significant AGN component within their modelled SEDs. Note that redshifts in the DEVILS samples are taken from a combination of spectroscopic, grism and photometric redshifts. The fraction of spectroscopic redshifts ranges from 90\,\% in D10 to 56\,\% in D02.

\begin{figure}
    \centering
    \includegraphics[width=0.9\columnwidth]{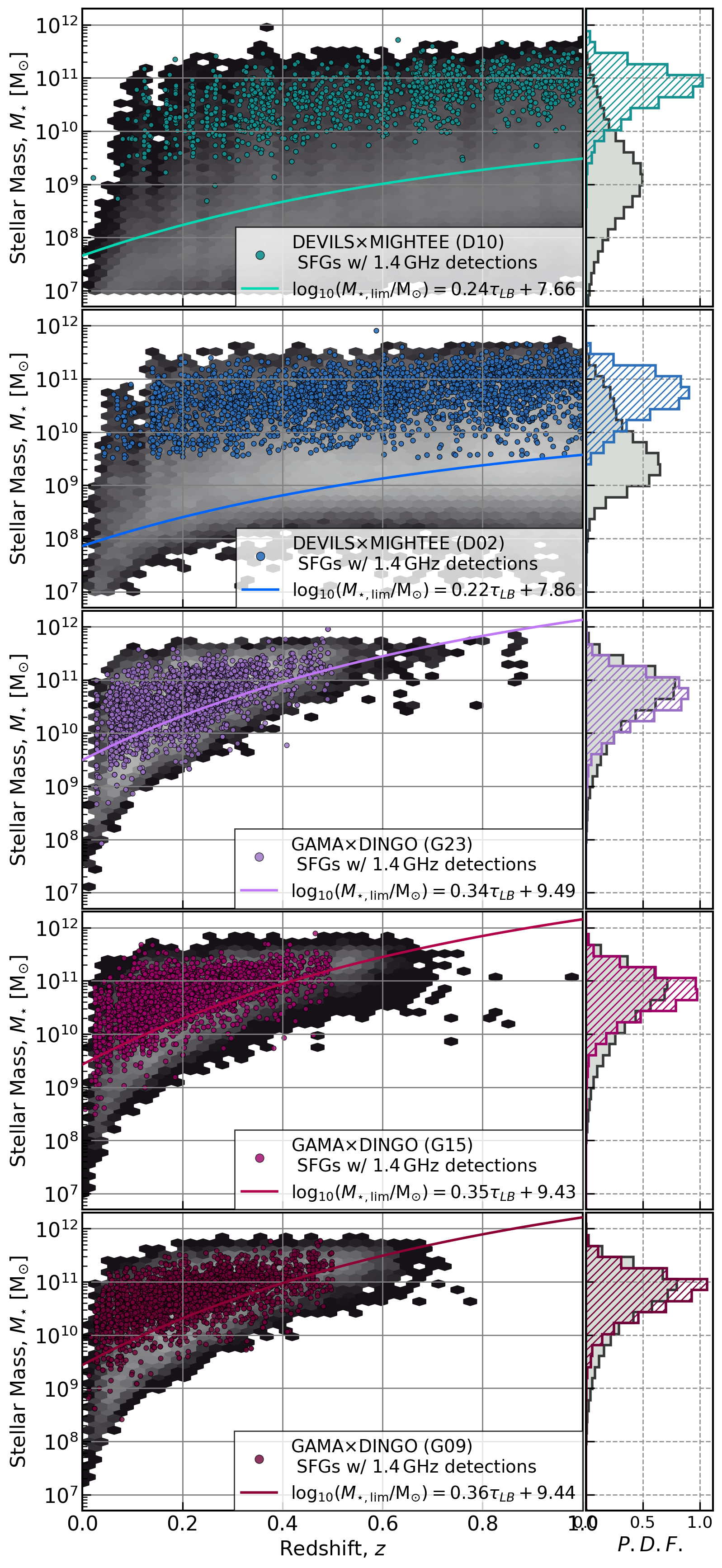}
    \caption{Stellar mass distribution plotted against redshift. The rows correspond to each of the fields with the COSMOS (D10) and XMM-LSS (D02) regions of the \DEVILSxMIGHTEE sample in the top panels followed by the \GAMAxDINGO samples in the G23, G15 and G09 fields. The grey scale in each panel represents the underlying number density of galaxies in the multi-wavelength catalogues (increasing logarithmically from black to white) and each coloured point indicates a radio source detection within 3$''$ of a star-forming galaxy (as per \citealt{Thorne2021}) and showing no signs of an AGN (see Section\,\ref{sec:data_and_sample} for details). Solid lines show the stellar mass completeness limits for each of the multi-wavelength surveys. Histograms on the right show the respective probability density functions for the full catalogues (grey) and the selected samples (hatched, coloured).}
    \label{fig:Mstar_vs_redshift}
\end{figure}

\subsection{Completeness in the matched samples}
\label{sec:sensitivities_in_samples}
Combining the \GAMAxDINGO and \DEVILSxMIGHTEE surveys in this way is, in some ways, similar to having two tiers of a ``wedding cake'' survey, a standard approach taken to provide a large number of galaxies at both low and high redshift. 

Figure\,\ref{fig:Mstar_vs_redshift} shows the distributions of stellar masses against redshift for each of the five multi-wavelength datasets overplotted by radio detections of SFGs in the respective radio continuum surveys. As in previous works \citep{Davies2018, Wright2018, Thorne2021}, we select volume-complete samples from each of the full datasets at each redshift. We estimate the mass completeness limits using the unattenuated, rest-frame $g-i$ colour distributions in intervals across the redshift range. At a given lookback time interval, we limit the samples to the lowest stellar mass ($M_\mathrm{lim}$) above which the sample is complete to the 90$^\mathrm{th}$ percentile of rest-frame $g-i$ colours. We replicate the approach of \citet{Thorne2021}, whereby this limit is defined by an equation with evolves linearly with lookback time; see e.g. their equation 3. The lines in Figure\,\ref{fig:Mstar_vs_redshift} (with corresponding equations given in each legend) show the stellar mass completeness cuts used in each field. The vast majority of MIGHTEE-detected sources are complete in stellar mass in the DEVILS observations, however, the same is not true for the DINGO sources.\\

\begin{figure}
    \centering
    \includegraphics[width=\columnwidth]{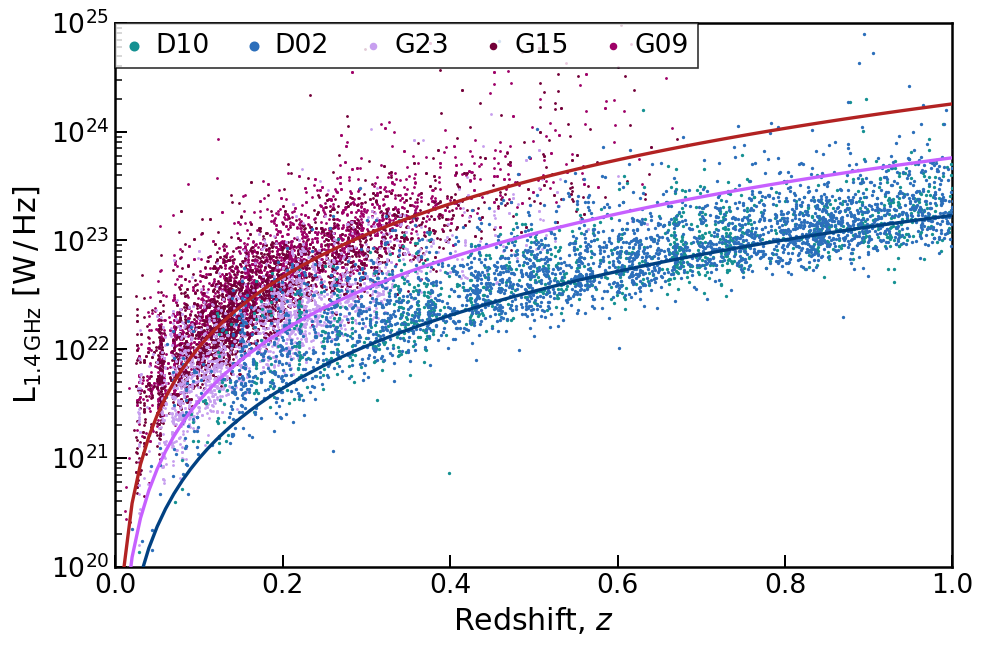}
    \caption{1.4\,GHz radio luminosity (assuming spectral index $\alpha=0.7$) against redshift for the cross-matched datasets used in this work. The points show the sample of star-forming galaxies above the mass completeness limits (see Figure\,\ref{fig:Mstar_vs_redshift}) for each field. The lines overplotted show the radio completeness flux density limits for DEVILS (blue, bottom), GAMA G23 (purple, middle) and GAMA G15 and G09 (red, top).}
    \label{fig:L1.4GHz_vs_z}
\end{figure}

Separately from the stellar mass completeness limits, we also consider whether each sample at a given redshift is complete in their radio continuum measurements, i.e. above what radio luminosity will an SFG with radio emission be detected in the radio continuum images. Figure\,\ref{fig:L1.4GHz_vs_z} shows the radio luminosity against redshift for the samples with the radio completeness limits overplotted. The limits not only differ between surveys but also between fields --- in particular due to deeper DINGO observations in the G23 field. Radio continuum observations are generally less sensitive to a given SFR than the most sensitive mid- and far-infrared filters, hence the threshold for an SFG entering the cross-matched sample is dictated by the depth of the radio continuum.

By injecting simulated radio sources into the MIGHTEE images and measuring their re-extracted fluxes, \citet{Hale2023} determined the flux density completeness limit to be 0.05\,mJy in MIGHTEE. We convert this 1.4\,GHz flux density limit to an SFR limit based on the infrared flux derived from Equation\,\ref{eqn:qTIR}, assuming a \qTIR value that is 1$\sigma$ (0.21\,dex) above the average for the entire sample ($\langle q_\mathrm{TIR}\rangle = 2.24$). Galaxies in the sample that fall below this SFR limit are excluded from the subsequent analyses. This ensures that at a given redshift, the \qTIR distribution is not significantly biased (within 1\,$\sigma$) against galaxies with high \qTIR, which may preferentially be detected at infrared wavelengths, but not at 1.4\,GHz.

We replicate the flux completeness limits from \citet{Hale2023} less rigorously in the DINGO fields by measuring the turnover of the radio luminosity number density as a function of redshift. In the range of flux densities used in this work, predicted models \citep[e.g. TRECS;][]{Bonaldi2019} and deep observations (e.g. from DEEP2, \citealt{Matthews2021}) expect radio source counts to increase with decreasing flux density. Based on this assumption, any turnover in the source counts seen at lower flux densities can likely be attributed to incompleteness. Our estimates effectively limit the DINGO detections of 1.4\,GHz flux densities to above 0.14\,mJy in G23 and 0.42\,mJy in both G15 and G09, which are then converted to an SFR limit in the same manner as for the MIGHTEE datasets. As validation, applying this approximation to the MIGHTEE datasets gives an almost identical flux density limit to that calculated by injecting mock sources as in \citet{Hale2023}.\\

The final stellar mass and luminosity complete samples result in 5,530 SFGs with no indication of a significant AGN component. This sample forms the basis for the remainder of the analysis presented in this paper.

\begin{figure*}
    \centering
    \includegraphics[width=0.8\textwidth]{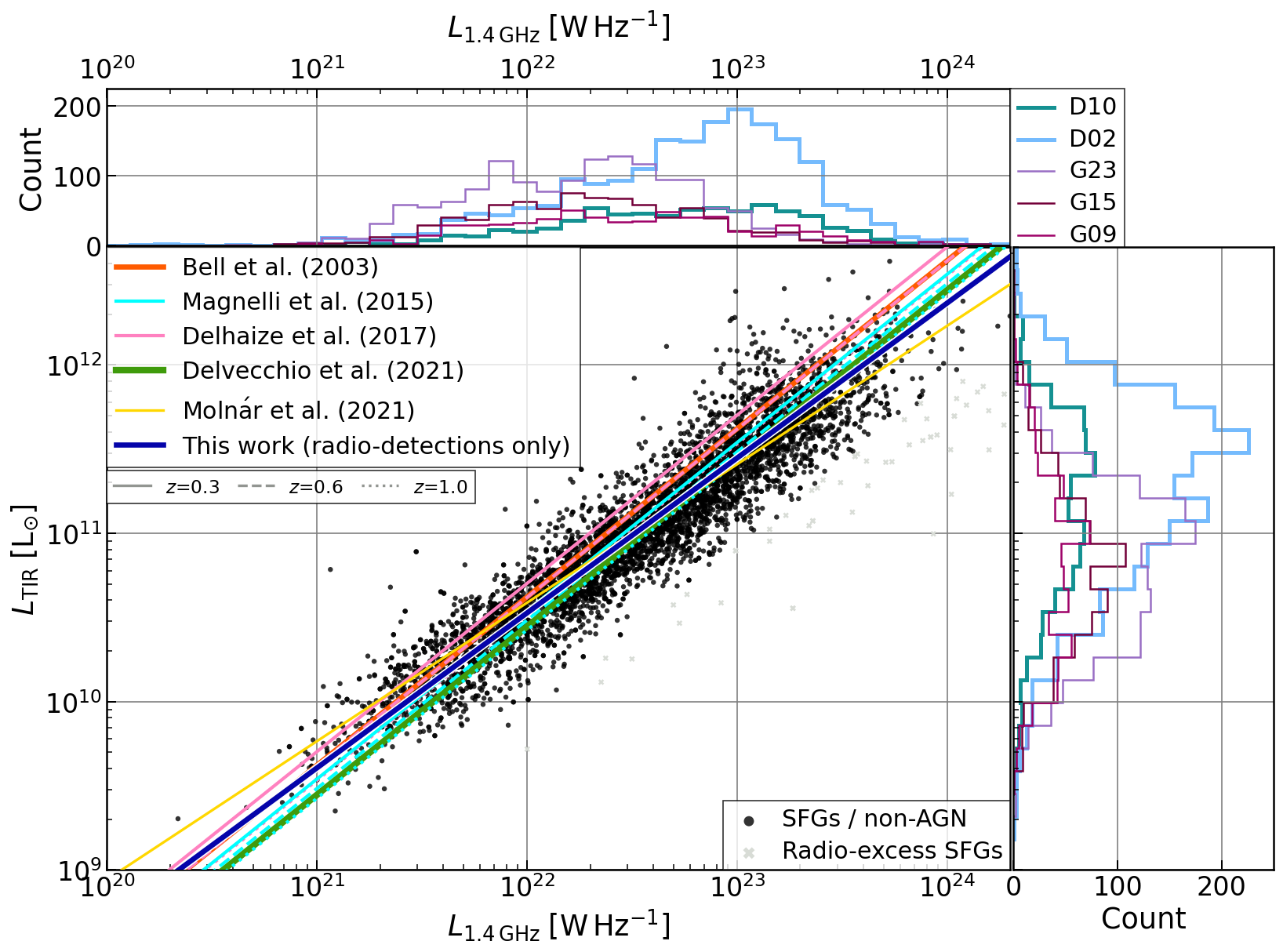}
    \caption{The IRRC of star-forming galaxies (black) presented as the total (8\,--\,1000\,$\mu$m) infrared luminosity against the 1.4\,GHz radio luminosity for all non-AGN radio sources cross-matched in \DEVILSxMIGHTEE D10 and D02 fields as well as the \GAMAxDINGO G23, G15 and G09 fields. Grey points indicate radio detections that are either flagged as hosting an AGN or non-star-forming ($<$0.5\,dex below the SFMS). The adjacent panels show histograms of \LTIR (right) and \Lradio (top) in each of these fields. As well as the best-fitting relation for this work in blue, various studies using different approaches for parameterising the IRRC are also shown, including seminal work from \citet{Bell2003}. \citet{Magnelli2015} use an \Mstar-selected sample with stacking of non-detections. \citet{Delhaize2017} use a joint infrared and radio-selected sample, accounting for non-detection with a survival analysis. \citet{Delvecchio2021} use a \Mstar-selected sample with stacking and simultaneously binned in both \Mstar and $z$. \citet{Molnar2021} use a matched infrared- and radio-selected sample at low redshift ($z < 0.2$). Solid, dashed and dotted lines represent redshifts of $z = $ 0.3, 0.6 and 1.0 for those that measure a significant redshift dependence in their parameterisations.}
    \label{fig:LTIR_vs_L1.4GHz_all_datasets}
\end{figure*}

\section{Results}
\label{sec:results}

\subsection{The \texorpdfstring{$L_\mathrm{TIR}$}{LTIR} - \texorpdfstring{$L_\mathrm{1.4GHz}$}{L1.4GHz} Relation}
\label{sec:LTIR_vs_L1.4GHz}
A key goal of this paper is to investigate the connection between the radio SFRs with more commonly used indicators that rely on emission at ultraviolet to infrared wavelengths. Historically, studies have presented this connection in the form of the infrared-radio correlation. Figure \ref{fig:LTIR_vs_L1.4GHz_all_datasets} shows the total infrared luminosity (\LTIR over 8\,--\,1000\,\microns) against \Lradio for the combination of all five cross-matched datasets used in this work.\\

We convert the observed radio flux densities from the radio continuum sources into rest-frame 1.4\,GHz radio luminosities ($L_\mathrm{1.4\,GHz}$) using the following equation:
\begin{equation}
    \frac{L_\mathrm{1.4GHz}}{\mathrm{W\,Hz^{-1}}} = \frac{4\pi}{(1+z)^{1-\alpha}} \left(\frac{3.086\xtimes10^{22}\,D_\mathrm{L}}{\mathrm{Mpc}}\right)^{2} \frac{1\xtimes10^{-26} S_\mathrm{1.4GHz}}{\mathrm{Jy}},
    \label{eqn:L1.4GHz}
\end{equation}

\noindent 
where $D_\mathrm{L}$ is the luminosity distance, $z$ is the redshift and $S_\mathrm{1.4\,GHz}$ is the integrated flux density with the factor of $1\times10^{-26}$ converting Jy to W\,m$^{-2}$\,Hz$^{-1}$. The redshifts are taken from the compilation of multi-wavelength catalogues described in \citet{Thorne2021} for the DEVILS samples and \citet{Driver2022} for the GAMA samples. We assume a spectral index of $\alpha=0.7$. A scatter of 0.35\,dex is observed in radio spectral indices \citep[e.g.][]{Smolcic2017b, An2021}, however, a constant universal value between 0.7\,--\,0.8 is often assumed as an average for the source population in a large sample accounting for both synchrotron and thermal free-free emission.

We measure \LTIR by extracting the infrared flux from the best-fitting SED model, which is modelled as a combination of a dust-attenuated stellar emission component, the re-emitted dust component and an additional AGN component. Note again that galaxies are removed from the samples if $f_\mathrm{AGN} > 0.1$ or show a radio luminosity in excess of $2\sigma$ ($\sim$0.42\,dex) from the global IRRC (see Section\,\ref{sec:selecting_SFGs}). The \LTIR is then calculated by integrating the SED over the rest-frame wavelength range of 8\,--\,1000\,\microns. As has been shown in many previous studies, the \LTIR-\Lradio relation in Figure\,\ref{fig:LTIR_vs_L1.4GHz_all_datasets} follows a roughly constant power-law trend over the approximately three orders of magnitude in radio luminosity shown here. The \GAMAxDINGO fields cover a lower redshift range than both DEVILS datasets, probing the IRRC to fainter radio luminosities and, hence, lower star formation activities. Using the multi-dimensional Markov Chain Monte Carlo (MCMC) fitting package \Hyperfit\footnote{http://hyperfit.icrar.org/} \citep{Robotham2015}, we fit a power-law relation to the IRRC for all datasets combined, which is given by:

\begin{equation}
    L_\mathrm{TIR}/L_\mathrm{\odot} = 10^{11.170\pm0.003}\,\left(\frac{L_\mathrm{1.4\,GHz} / \mathrm{W\,Hz^{-1}}}{5\times10^{22}}\right)^{0.921\pm0.004},
    \label{eqn:IRRC_hyperfit}
\end{equation}

\noindent
with an orthogonal scatter of 0.137\,dex. We find that the IRRC as measured by combining the SKAO precursor telescopes has a lower overall scatter to those previously obtained using vastly different datasets, which range from 0.16\,--\,0.3\,dex \citep{Yun2001, Murphy2008, Molnar2021}. The sub-linear behaviour (slope < 1) in the IRRC has also been noted in several recent papers \citep{Hodge2008, LoFaro2015, Davies2017, Gurkan2018, Molnar2021, Delvecchio2021}.

\citet{Bell2003} quantified the IRRC by a single median value of \qTIR = $2.64\,\pm\,0.02$ and a corresponding scatter of 0.26\,dex. However, more recent studies have found \qTIR to vary with redshift \citep[e.g.][]{Magnelli2015, Delhaize2017, Basu2015}, with stellar mass \citep[e.g.][]{Delvecchio2021} and with spectral index, $\alpha$ \citep[e.g.][]{An2021}. \citet{Magnelli2015}\footnote{Note, as per \citet{Magnelli2015}, we scale their relation by a factor of log10(1.91) to account for the conversion between far infrared (40\,--\,120\microns) to the total infrared defined here.} parameterised this relation at far-infrared wavelengths with a stellar mass-selected sample below $z\,<\,2$, showing that \qFIR evolves with redshift according to $q_\mathrm{FIR}(z) = (2.35\,\pm\,0.08)\left(1 + z\right)^{-0.12\,\pm\,0.04}$. \citet{Delhaize2017} used a sample of infrared-detected (in \textit{Herschel}) and radio-detected (3\,GHz from VLA) sources to find a redshift evolution of the form $q_\mathrm{TIR}(z) = (2.88\,\pm\,0.03)(1 + z)^{-0.19\,\pm\,0.01}$. In Figure\,\ref{fig:LTIR_vs_L1.4GHz_all_datasets}, we represent these observed redshift evolutions with solid, dashed and dotted lines denoting $z =$ 0.3, 0.6 and 1.0, respectively. More recently, \citet{Delvecchio2021} addressed this trend through a bivariate analysis by simultaneously studying the \qTIR with stellar mass and redshift, concluding that increasing \Mstar was the primary cause for a lower \qTIR, whereas redshift has only a secondary impact.

With regards to AGN contamination, \citet{Magnelli2015} and \citet{Delhaize2017} used median stacking and sigma clipping (at $3\,\sigma$) to mitigate the impact of AGN with excess radio emission. \citet{Delvecchio2021} implemented a more rigorous exclusion of radio-excess sources, making a cut below a threshold of 2$\sigma$ from the peak of the \qTIR for SFGs at a given redshift and stellar mass. This assumes that the intrinsic scatter of \qTIR for SFGs is symmetric about its peak and that radio-excess AGN are not the dominant population (however, note figure 17 of \citealt{Thorne2022b}). At high stellar masses, contamination from radio-quiet AGN can be significant when only using sigma clipping alone. \citet{Delvecchio2021} show that for their highest stellar mass bins (\Mstar\,$\geqslant\, 10^{10.5}$\,\Msun), $\sim$80\,per cent of AGN are found below a threshold of 2$\sigma$ from the peak \qTIR, but with a highly complete sample of SFGs found above this cut.

Following a similar strategy to \citet{Delvecchio2021}, we additionally remove sources from our SFG samples that show a large radio excess to remove potential outliers caused by incorrect matching of optical sources with bright AGN features or sidelobe artefacts. We define our cut at 2$\sigma$ below the average \qTIR, where the vertical dispersion in the relation has been measured to be $\sigma = 0.21$\,dex. Removing the additional AGN cuts from \citet{Whittam2022} applied to the D10 field only marginally increases the orthogonal scatter, which suggests that the low overall scatter in the D10 relation is due to the high-quality photometry available in COSMOS. The results of this paper are unchanged with these additional cuts removed. \\

\begin{figure}
    \centering
    \includegraphics[width=0.48\textwidth]{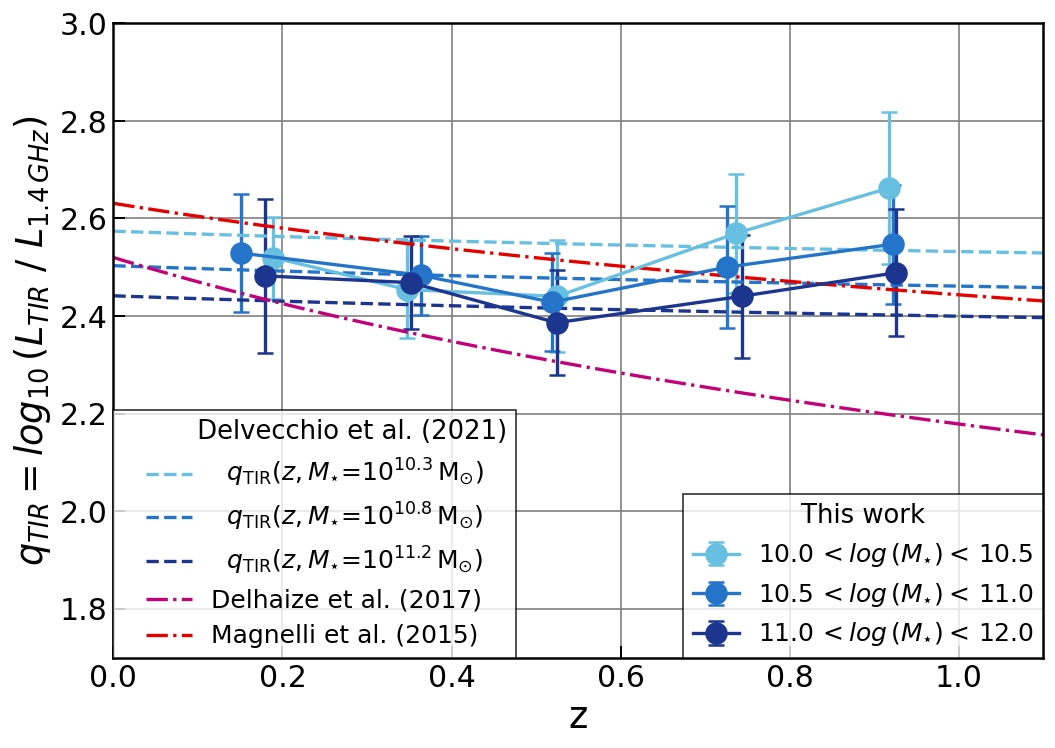}
    \caption{The evolution of the median \qTIR over redshift separated into bins of \Mstar shown as solid lines with error bars showing the interquartile range of points within each bin. The dashed lines show literature values for \qTIR$(z)$ from \citet{Delhaize2017} (magenta) and \citet{Magnelli2015} (red). The multivariate \qTIR$(z,\, M_{\star})$ expression of \citet{Delvecchio2021} is shown as coloured lines at the mean \Mstar values in each of our stellar mass bins. After applying \Mstar and \Lradio completeness cuts, only bins with greater than five points are shown.}
    \label{fig:qTIR_vs_z}
\end{figure}

\begin{figure*}
    \centering
    \includegraphics[width=0.8\textwidth]{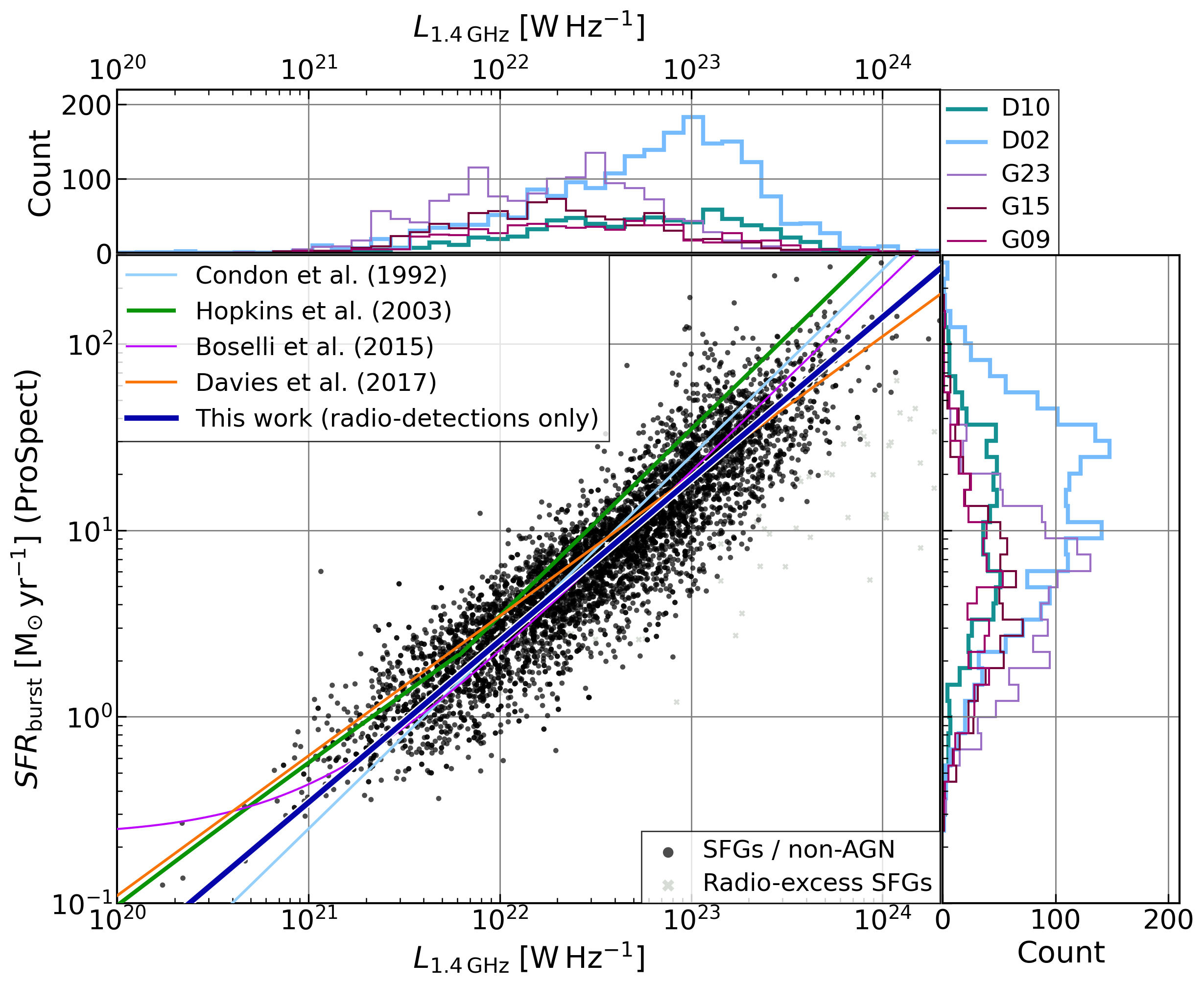}
    \caption{The SFR-\Lradio relation of all star-forming, non-AGN galaxies (black) in all of the cross-matched multi-wavelength and radio continuum fields used in this work. Grey points indicate cross-matched radio detections flagged as hosting an AGN or non-star-forming; defined as being $<$0.5\,dex below the SFMS. Adjacent panels show histograms of SFR (right) and \Lradio (top) for the sample (black points) separated by the field. The dark blue line represents a power-law fit to the combination of datasets used here. Several 1.4\,GHz calibrations for SFR from the literature (including \citealt{Condon1992, Hopkins2003, Boselli2015, Davies2017}) have also been overplotted for reference. Where necessary, these SFRs have been rescaled to a Chabrier IMF.}
    \label{fig:SFR_vs_L1.4GHz}
\end{figure*}

\subsection{Correlation of \texorpdfstring{\qTIR}{qTIR} with Galaxy Properties}
\label{sec:qTIR_by_redshift}

It has become commonplace to investigate how \qTIR varies as a function of various host galaxy properties. In Figure\,\ref{fig:qTIR_vs_z}, we show the redshift evolution of the average \qTIR of 1.4\,GHz-detected SFGs further separated into three bins of stellar mass. The stellar mass and radio luminosity completeness cuts allow us to investigate the redshift evolution for stellar masses above \Mstar$=10^{10}$\,\Msun --- below this, the sample is highly stellar mass incomplete within the GAMA samples. The points show the median \qTIR with error bars indicating the spread of data as the interquartile range.

Within each stellar mass bin, we see no consistent decrease in \qTIR over the entire redshift range --- in contrast with previous works such as \citet{Magnelli2015, Delhaize2017} (dash-dotted lines). The fact that the trend between \qTIR and $z$ is not monotonically decreasing implies that the sub-unity slope of the \LTIR--\Lradio relation observed in Figure\,\ref{fig:LTIR_vs_L1.4GHz_all_datasets} is likely not due to a redshift evolution of the zero point of an otherwise unity \LTIR--\Lradio relation. This implies that the ratio between the infrared and radio emission is not constant with luminosity, and therefore differs between galaxies of different properties; for example with star formation activity, as has been discussed in \citet{Molnar2021}.\\

On the other hand, there is a slight indication that \qTIR decreases with increasing stellar mass, differing by 0.08\,dex between $\langle\log(M_{\star})\rangle = 10.2$\,\Msun\,--\,11.2\,\Msun. It should be noted that the stellar mass trend here is only present at redshifts above $z > 0.5$ and the spread in \qTIR values within a given bin of redshift and stellar mass is typically larger than the separation across redshift bins. Such trend with stellar mass was also observed in \citet{Delvecchio2021}, shown as blue dashed lines for similar bins in redshift and stellar mass\footnote{As \Mstar estimates from \ProSpect are 0.2\,dex higher on average than \textsc{Magphys} \citep{Thorne2021}, we homogenise their values by scaling the stellar masses used for these \qTIR comparisons.}. The \qTIR measurements presented here are broadly consistent with those presented in \citet{Delvecchio2021} at high mass, however, in our lower \Mstar bin, the \qTIR values are, on average, $\sim0.1$\,dex lower than the \citet{Delvecchio2021} relations. This difference may be due to the fact that our sample traces only radio detections, meaning that in regimes where our sample becomes increasingly incomplete, we sample only the brightest radio sources, which may lead to a lower average \qTIR as per the sub-unity trend in the IRRC. Further to the sample selection, differences in the SED fitting process (e.g. inclusion of an AGN component) can lead to a reduction in the integrated infrared luminosity.\\

Expressing \qTIR in terms of the proportionality observed in the IRRC gives \qTIR = $\log_{10}\left(L_\mathrm{TIR} / L_\mathrm{1.4\,GHz}\right) = \log_{10}\left(L_\mathrm{1.4\,GHz}\,^{\gamma} / L_\mathrm{1.4\,GHz}\right)$, i.e., \qTIR $\propto (\gamma - 1)\times\,\log_{10}(L_\mathrm{1.4\,GHz})$. Taking the value of the slope $\gamma \sim 0.9$ from Equation\,\ref{eqn:IRRC_hyperfit}, for every order of magnitude increase in \Lradio, \qTIR decreases by $\sim$0.1\,dex. Most galaxy properties scale with the stellar mass of a galaxy; for instance, for a $\sim$1\,dex change in the stellar masses, the infrared luminosity will on average increase by $\sim$0.8\,dex, which would correspond to \qTIR decreasing by $\sim$\,0.08\,dex assuming the IRRC power-law slope above. In Figure\,\ref{fig:qTIR_vs_z}, the lowest and highest stellar mass bins differ by $\sim1$\,dex in their median \Mstar and the average \qTIR offset of 0.08\,dex closely matches the expectation above. Note, however, that due to the fact that the normalisation of the SFMS increases by $\sim1$\,dex from $z=0$ to $z=1$, one would also expect that the average \qTIR of galaxies decreases by $\sim0.1$\,dex over this redshift range given the sub-unity slope of the \LTIR--\Lradio relation.\\

\subsection{SFR-\texorpdfstring{\Lradio}{L1.4GHz} Relation}
\label{sec:SFR_vs_L1.4GHz}
Whilst infrared luminosity has historically been used as a proxy for star formation rates in galaxies, the relation may only hold for massive, dusty star-forming galaxies. Thus to obtain reliable SFR estimates in less dusty systems, often the infrared dust emission is complimented with ultraviolet photometry \citep[UV+TIR; see e.g., ][]{Brown2014, Davies2017}. In low metallicity and high redshift galaxies, the UV can contribute as much to the total SFR as the IR alone \citep{Whitaker2017}. In this work, we use SFR values derived from FUV\,--\,FIR SED fits, which provide a physically-motivated estimate for the recent SFR and better control over associated errors \citep{Davies2016}.

\begin{table*}
    \centering
    \caption{The resulting best-fitting parameters from running \Hyperfit on both the \LTIR-\Lradio and SFR-\Lradio relations expressed as power-laws of form: $\log_{10}(y) = \gamma\times\log_{10}\left(L_\mathrm{1.4\,GHz}\,/\,5\times10^{22}\right) + \beta$, and with an orthogonal scatter, $\sigma_{\perp}$, fit as an additional parameter. Additionally, we include the average \qTIR value for SFGs in each sample. \LTIR is measured in units of $\mathrm{L_{\odot}}$, $SFR$ in units of \MSunperyear and \Lradio in units \WperHz. The errors quoted are the standard fitting errors on the parameters.}
    \begin{tabular}{ccccccccc} \hline
        Sample & \multicolumn{3}{c}{\LTIR\,-\,\Lradio} & \aveqTIR & \multicolumn{3}{c}{SFR\,-\,\Lradio} \\
         & $\gamma$ & $\beta$ & $\sigma_{\perp}$ & & $\gamma$ & $\beta$ & $\sigma_{\perp}$ &  \\ \hline
        D10 & 0.911$\pm$0.01 & 11.130$\pm$0.005 & 0.106$\pm$0.003 & 2.435 & 0.84$\pm$0.01 & 0.977$\pm$0.007 & 0.142$\pm$0.004 \\
        D02 & 0.969$\pm$0.008 & 11.192$\pm$0.004 & 0.147$\pm$0.002 & 2.487 & 0.90$\pm$0.01 & 1.046$\pm$0.005 & 0.165$\pm$0.003 \\
        G23 & 0.865$\pm$0.008 & 11.205$\pm$0.005 & 0.108$\pm$0.002 & 2.571 & 0.81$\pm$0.01 & 1.033$\pm$0.007 & 0.151$\pm$0.003 \\
        G15 & 0.898$\pm$0.011 & 11.158$\pm$0.008 & 0.127$\pm$0.003 & 2.507 & 0.85$\pm$0.01 & 0.99$\pm$0.01 & 0.153$\pm$0.004 \\
        G09 & 0.909$\pm$0.012 & 10.998$\pm$0.007 & 0.117$\pm$0.003 & 2.330 & 0.84$\pm$0.01 & 0.848$\pm$0.008 & 0.131$\pm$0.004 \\
        All & 0.921$\pm$0.004 & 11.170$\pm$0.003 & 0.137$\pm$0.001 & 2.490 & 0.868$\pm$0.005 & 1.014$\pm$0.003 & 0.161$\pm$0.002 \\
        \hline
    \end{tabular}
    \label{tab:LTIR_vs_L1.4GHz_params}
\end{table*}

Figure \ref{fig:SFR_vs_L1.4GHz} presents the SFR integrated over the last 100\,Myr (\SFRburst) from \ProSpect against the 1.4\,GHz radio luminosity for star-forming galaxies from all of the datasets. Points are similarly coloured as for Figure\,\ref{fig:LTIR_vs_L1.4GHz_all_datasets} with black points denoting SFGs that show no presence of an AGN. The blue line shows the best-fitting power-law relation to these black points only, given by the equation:

\begin{equation}
    \begin{split}
    \frac{SFR}{\mathrm{M_{\odot}\,yr^{-1}}} = 10^{1.014\pm0.003}\cdot\left(\frac{L_\mathrm{1.4\,GHz}}{5\times10^{22}\,\mathrm{W\,Hz^{-1}}}\right)^{0.868\pm0.005}.
    \end{split}
    \label{eqn:SFR_vs_Lradio}
\end{equation}

\noindent
As was shown in \citet{Molnar2021}, fitting a power-law relation to this SFR calibration takes into account the dependence of \qTIR on \Lradio, as was also done in \citet{Davies2017}. The best-fitting relation has an orthogonal scatter of 0.16\,dex, which is slightly larger than for the fields fit independently --- particularly D10. The larger scatter in the D02 field is unsurprising as despite spanning a similar redshift range to D10, it contains poorer quality multi-wavelength photometry and, hence, less constrained SED parameters. This is particularly true for the NIR photometry that for D02 comes from VIDEO\footnote{VIDEO: VISTA Deep Extragalactic Observations survey.} \citep{Jarvis2013}, which is shallower in depth than the UltraVISTA \citep{McCracken2012} data used in D10. There is also a greater proportion of radio-excess outlier sources in D02 compared to D10, due to the lack of detailed AGN classifications, which will be made available for other fields in later data releases of MIGHTEE. The SFR-\Lradio relation shows a higher orthogonal scatter than the corresponding \LTIR-\Lradio relation for each field.

There is relatively good agreement between this work and the radio SFR calibrations from the literature shown in Figure\,\ref{fig:SFR_vs_L1.4GHz}.
A \emph{break} in the linearity of the SFR-\Lradio relation has been suggested in previous works, requiring calibrations such as those from \citet{Boselli2015} and \citet{Hopkins2003} to use either a non-linear (in logarithmic space) or piece-wise expression. A difference in slope may arise when averaging over different populations of galaxies where the link between the thermal (i.e. FUV\,--\,FIR) and non-thermal emission (i.e. \Lradio) differs between galaxies; e.g. at low luminosities \citep{Molnar2021} or at high redshift \citep{Delhaize2017}. Our data does not show a significant non-linearity at low \Lradio, however, at $L_\mathrm{1.4\,GHz} \gtrsim 10^{23}\,\mathrm{W\,Hz^{-1}}$, we observe higher SFRs on average than the overall fitted relation. Thus at high \Lradio our data matches more closely with the steeper slopes measured in \citet{Condon1992}, \citet{Boselli2015} and \citet{Hopkins2003}. Furthermore, our sample consists of multiple fields across two separate multi-wavelength surveys that, despite having been reduced with a consistent set of software and techniques, will differ in the quality of their photometric measurements. The DEVILS D10 and D02 samples extend up to $z\sim1$, where galaxies are evolving under different conditions than in the local Universe. One would therefore expect larger variations in galaxy SFHs, which may present as offsets between the thermal and non-thermal emission mechanisms that probe star formation on different timescales \citep{Arango-Toro2023}. In the following sections, we consider the impact of a galaxy's star formation history on SFRs derived from radio observations.

\section{Analysis}
\label{sec:analysis}
In the following sections, we build upon the results of the SFR-\Lradio relation above to investigate the impact of timescales in SFR calibrations to address future usage of radio continuum emission as a universal star formation rate indicator.

\subsection{Measuring variability of star formation histories}
\label{sec:defining_SFHs}
The infrared and SED-derived SFRs are sensitive to relatively recent star formation on timescales of order of the lifetimes of the OB stars responsible for emitting radiation at these wavelengths. On the other hand, subsequent synchrotron emission triggered by the supernovae occurring in the most massive (\Mstar$\gtrsim8$\,\Msun) of these stars is delayed first by the lifespans of these stars and secondly by the fact that cosmic ray electrons must then be accelerated to relativistic speeds \citep{Roussel2003} on timescales that span much longer than the time taken for the responsible supernovae to fade entirely \citep{Pooley1969, Ilovaisky1972}.

We explore this discrepancy in timescales by leveraging the wealth of measured galaxy properties from the DEVILS and GAMA multi-wavelength catalogues, particularly the parameterised star formation histories (SFH) modelled from the \ProSpect SED fits. We define a metric for measuring the recent change in star formation rate as the net difference between the most recent star formation rate ($\mathrm{SFR}(t_\mathrm{LB}=0)$) and the SFR measured at a lookback time 200\,Myr prior
(hereafter \DeltaSFRMyr). We quantify the change in SFR as:

\begin{equation}
    \Delta SFR_\mathrm{200\,Myr} = SFR(t_\mathrm{LB}=0) - SFR(t_\mathrm{LB}=200\,\mathrm{Myr}),
    \label{eqn:deltaSFR}
\end{equation}

\noindent
In this parameterisation, a positive (negative) \DeltaSFRMyr indicates an increase (decline) in a galaxy's SFR since the previous epoch. The timescale of 200\,Myr is chosen as it is close to the finest time difference that \ProSpect can robustly quantify, meaning most SFHs within this short timescale are approximately linear. Longer timescales introduce situations where a given SFH can rise and fall to a similar $SFR$ value, incorrectly classifying an SFH that has recently declined as constant over time. Additionally, the $\Delta SFR$ values for galaxies younger than the timescale probed are difficult to define. Note, however, that using longer timescales up to \DeltaSFRGyr does not impact our results as most galaxies have fairly consistent SFH slopes between 200\,Myr and 1\,Gyr and so only a small fraction would give misleading SFH slopes.

\begin{figure}
    \centering
    \includegraphics[width=\columnwidth]{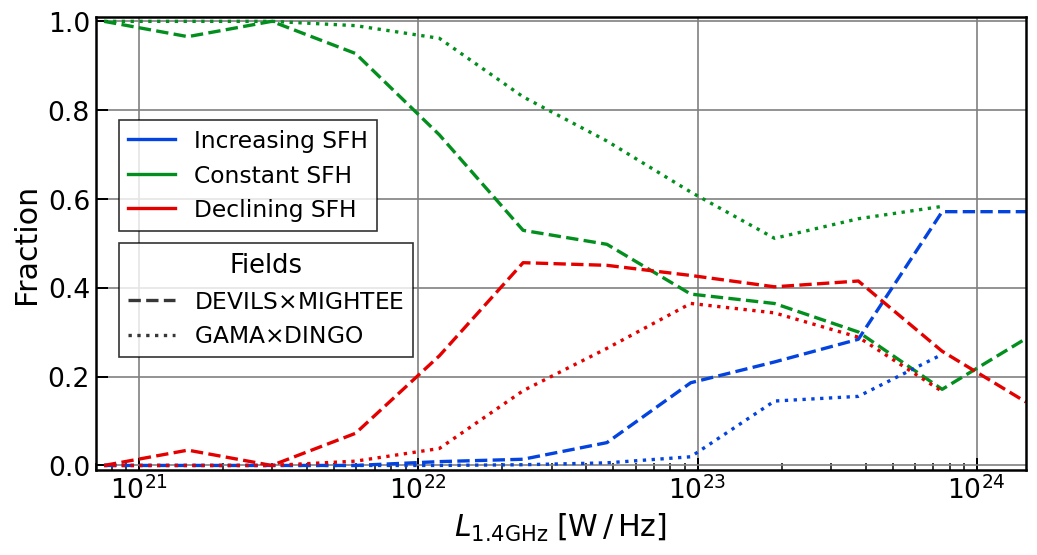}
    \includegraphics[width=\columnwidth]{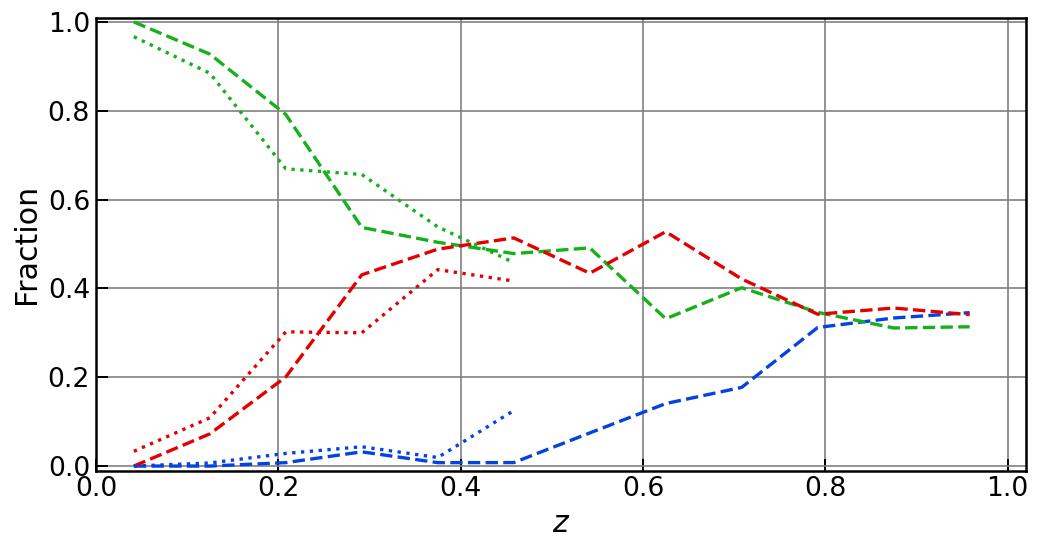}
    \caption{Fraction of SFHs classified as increasing (\DeltaSFRMyr $\geqslant$ 0.5\,\MSunperyear), decreasing (\DeltaSFRMyr $\leqslant$ -1.0\,\MSunperyear) and constant (-1.0 $<$ \DeltaSFRMyr $<$ 0.5\,\MSunperyear) against 1.4GHz luminosity (top) and redshift (bottom). The dashed and dotted lines correspond to the DEVILS and GAMA surveys, respectively.}
    \label{fig:SFH_fractions_vs_lum}
\end{figure}

This opens up a novel avenue for studying the evolution of galaxies that --- in addition to measuring the current rate of star formation --- can also distinguish between galaxies that are in the process of quenching or undergoing a starburst from those that have been quenched several billions of years ago. Other studies \citep[e.g.][]{Martin2017, deSa-Freitas2022} have noted similar trends using analogous quantities, such as a star formation \textit{acceleration}, which they define as the difference in star formation activity (as probed by $NUV-r$ colour) divided by the change in the time between two epochs.\\

The distribution of \DeltaSFRMyr for the DEVILS and GAMA surveys both peak at a constant SFH (\DeltaSFRMyr$=0$) and are skewed towards a net decrease in their recent SFH. Figure\,\ref{fig:SFH_fractions_vs_lum} shows the fraction of increasing, constant and declining SFHs as a function of 1.4\,GHz luminosity and redshift for both pairs of surveys. There are fractionally fewer SFGs with an increasing SFH in the GAMA datasets ($\sim1$\%) than in both the D10 (6\,\%) and D02 (11\,\%) regions, which is due to DEVILS extending to higher redshifts. This reflects the fact that at earlier epochs of the Universe, galaxies exhibited more bursty star formation \citep{Guo2016}, mergers were occurring more frequently \citep{Robotham2014, Keenan2014} and cold gas reservoirs were larger and more accessible for star formation \citep{Oteo2017}. This trend is also commensurate with the overall decline in the cosmic star formation history of the Universe which peaked around $z\sim2$ \citep{Madau&Dickinson2014, Driver2018, Bellstedt2020b}. It is worth emphasising that where the two surveys overlap below $z<0.5$, they share an almost identical distribution of SFHs as a function of redshift.\\

\begin{figure}
    \centering
    \includegraphics[width=\columnwidth]{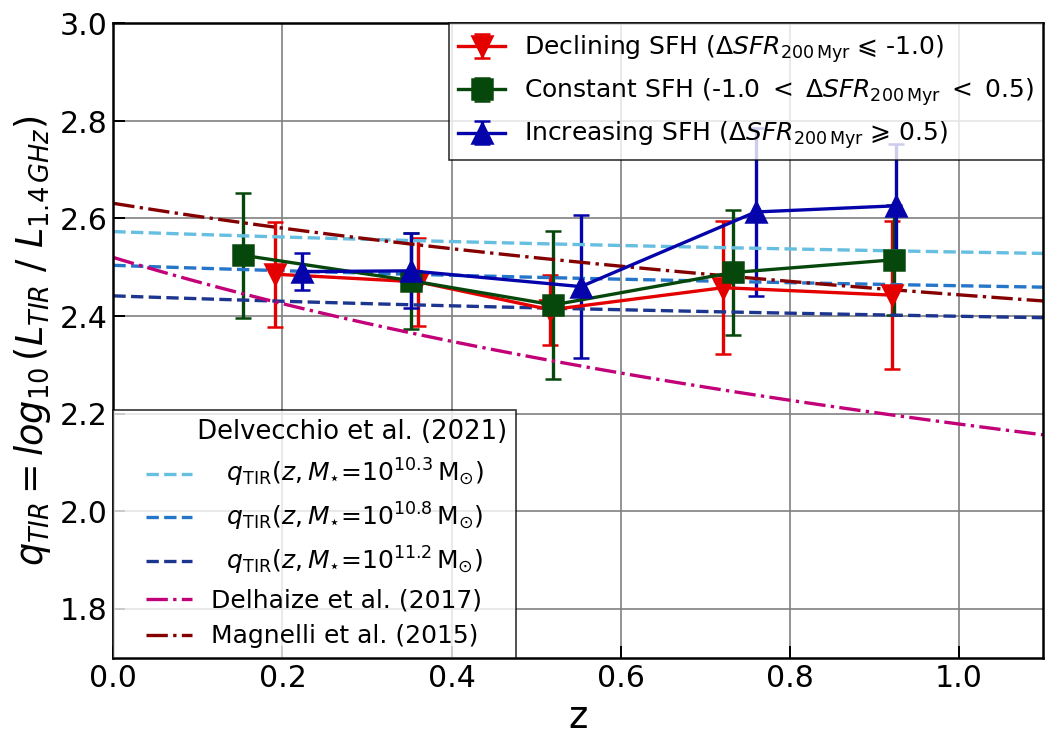}
    \caption{As per Figure\,\ref{fig:qTIR_vs_z}, except separated into three bins on \DeltaSFRMyr representing declining (red), constant (green) and increasing (blue) star formation histories. At higher redshifts, greater separation is seen between average \qTIR values than between bins of stellar mass. Only bins with greater than five data points are shown.}
    \label{fig:qTIR_vs_z_by_DeltaSFR}
\end{figure}

Figure\,\ref{fig:qTIR_vs_z_by_DeltaSFR} shows the redshift evolution of \qTIR separated into bins of \DeltaSFRMyr corresponding to increasing, constant and declining SFHs. Within each of the three bins, there is marginal redshift evolution seen by a gradual decrease at $z < 0.5$ followed by an increase out to $z=1$) This trend is consistent with separating into bins of stellar mass as in Figure\,\ref{fig:qTIR_vs_z}. However, galaxies with an increasing SFH have higher \qTIR values on average compared to a constant SFH. Galaxies with declining SFHs show only a marginally lower \qTIR in the highest redshift bin. This result indicates that the SFH of a galaxy may have an impact on the ratio between the thermal IR emission and non-thermal synchrotron. As well as increasing and declining SFHs becoming more frequent towards higher redshift (e.g. Figure\,\ref{fig:SFH_fractions_vs_lum}), the relative differences in SFR (i.e. \DeltaSFRMyr) also become much larger. Hence, this might explain why the impact of SFHs on \qTIR only becomes apparent above $z \gtrsim 0.5$. Below this redshift, the gradual decrease, which is commensurate with previous studies \citep[][]{Magnelli2015, Delhaize2017} --- and to a lesser extent \citet{Delvecchio2021} --- may instead be driven by the redshift evolution of the SFMS normalisation in combination with an IRRC with a sub-unity slope (see Section\,\ref{sec:qTIR_by_redshift}).

\begin{figure*}
    \centering
    \includegraphics[width=0.495\textwidth]{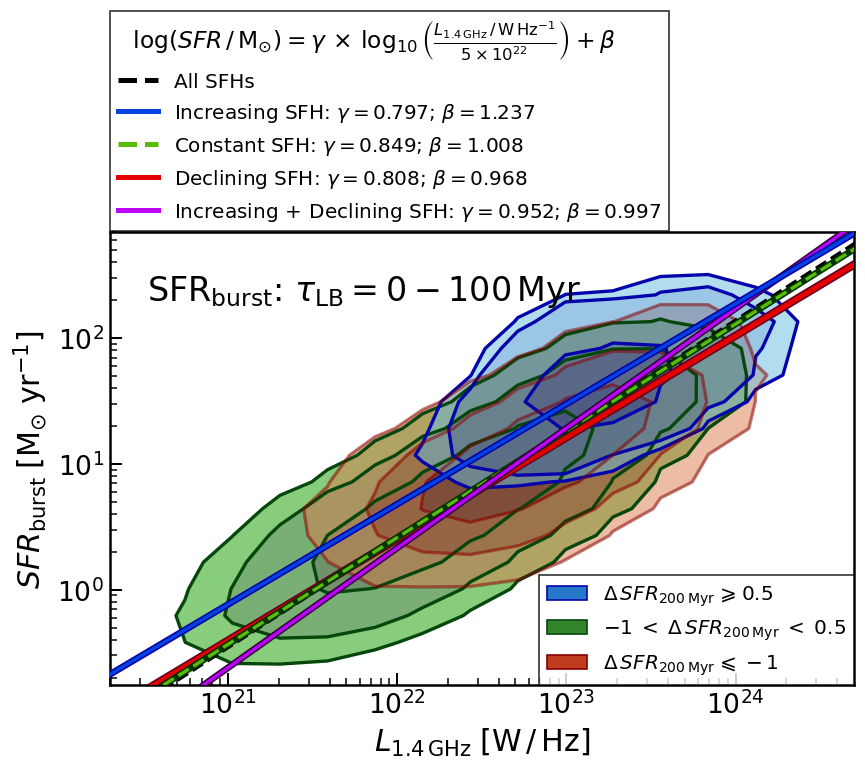}
    \includegraphics[width=0.495\textwidth]{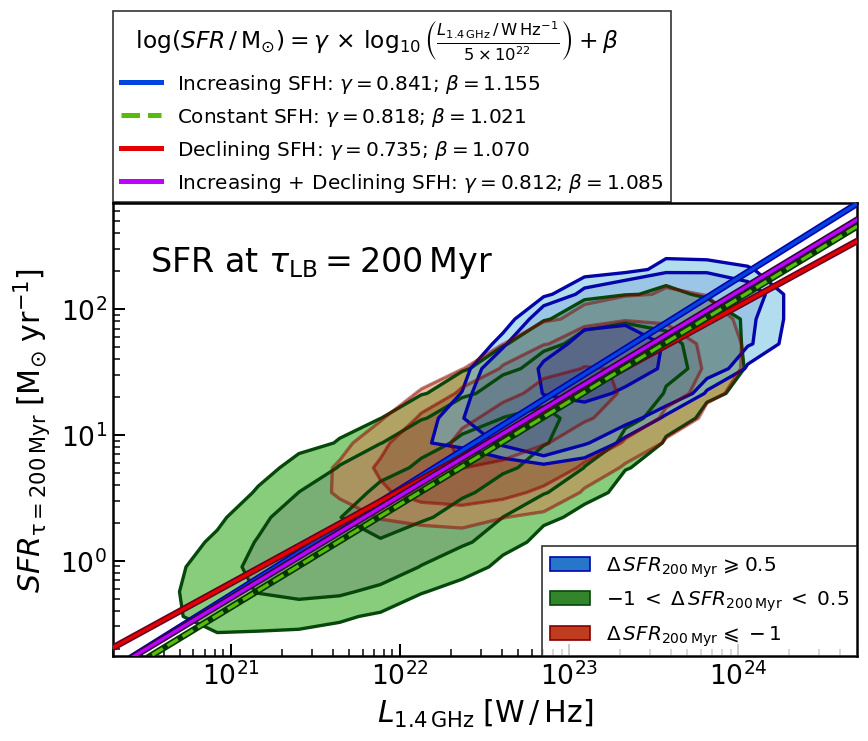}
    \caption{The SFR-\Lradio relation for SFR averaged over the last 100\,Myr (left) and a previous lookback time of $\tau_\mathrm{LB}=200$\,Myr (right). The sample has been separated into three populations based on \DeltaSFRMyr with green representing relatively constant SFHs, whereas blue and red show the extreme ends of the increasing and declining SFH distributions, respectively. The contour levels enclose percentiles of 68\%, 95.5\% and 99.5\% for each SFH subset. Solid lines show the best-fitting power-law relations with fitting parameters shown in the legend. The purple line denotes the combined set of increasing and declining SFHs, highlighting that the non-linear behaviour in the SFR-\Lradio could be driven by differences in SFH.}
    \label{fig:SFR_vs_L1.4GHz_SFH_subsets}
\end{figure*}

\subsection{Impact of timescales on the SFR-\texorpdfstring{\Lradio}{L1.4GHz} relation}
\label{sec:impact_of_sfr_timescales}
Figure\,\ref{fig:SFR_vs_L1.4GHz_SFH_subsets} shows the SFR-\Lradio relation for the combined datasets, however, with galaxies separated into populations based on the change in their star formation histories over the last 200\,Myr, \DeltaSFRMyr. Galaxies with an increasing SFH deviate the most from the power-law fit of all galaxies given in Figure\,\ref{fig:SFR_vs_L1.4GHz} and are, on average, situated 0.23\,dex above the relation for the constant SFH population. Those with declining SFHs, on the other hand, have SFRs that are very similar to constant SFHs --- an offset is only notable in the most extremely declining SFHs (i.e. $\Delta SFR_\mathrm{200\,Myr} < -3.0$\,\MSunperyear), which are few in number. The purple line represents the result of running \Hyperfit on the subsets of galaxies with increasing and declining SFHs. This combination results in the steepest slope in the SFR-\Lradio relation with a value that is closer to unity, $\gamma=0.95\pm0.03$. The non-linear aspect that has been noted in some SFR-\Lradio calibrations \citep[e.g.,][]{Hopkins2003, Boselli2015} may be associated with a larger fraction of increasing and declining SFHs towards brighter radio luminosities.

The right panel of Figure\,\ref{fig:SFR_vs_L1.4GHz_SFH_subsets} adjusts the SFR of each source from integrated SFR in the last 100\,Myr (\SFRburst) to the instantaneously measured SFR at a lookback time 200\,Myr prior in the modelled SFHs. Sources with increasing SFHs will by definition decrease in SFR with lookback time and vice versa for declining SFHs. As expected, the resulting best-fitting relation for galaxies with constant SFH remains mostly unchanged between these different epochs. Collectively, the increasing and declining SFHs regress to a relation that is much closer to that measured for constantly evolving galaxies. Figure\,\ref{fig:SFH_fractions_vs_lum} shows that towards higher redshifts, increasing and declining SFHs are more common, which is due to SFHs varying by greater amounts over a given timescale (e.g. \citealt{Guo2016}; Davies et al. submitted, MNRAS). Therefore the SFRs derived from ultraviolet--infrared emission and those measured at 1.4\,GHz will likely show greater differences as is seen by the larger separation of \qTIR values at higher redshifts in Figure\,\ref{fig:qTIR_vs_z_by_DeltaSFR}.

\subsection{Scatter in the \texorpdfstring{SFR}{SFR}-\texorpdfstring{\Lradio}{L1.4GHz} relation over time}
\label{sec:SFR_vs_L1.4GHz_over_time}

\begin{figure}
    \centering
    \includegraphics[width=\columnwidth]{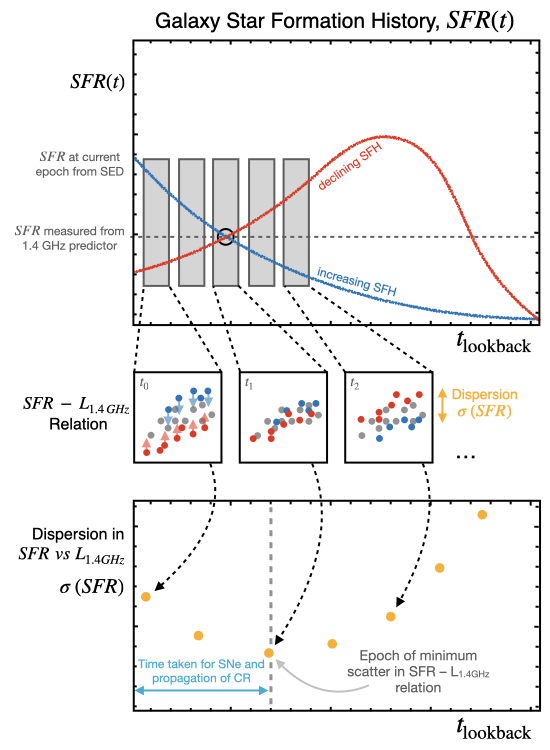}
    \caption{Infographic showing the procedure for estimating the relevant epoch at which the SFR determined from stellar emission best matches that determined from the $L_\mathrm{1.4\,GHz}$. The top panel shows a schematic of two star formation histories: one for a galaxy declining in its star formation history (red) and the other where its star formation is increasing (blue) with time. The middle panels represent several renditions of the SFR-\Lradio relation, except for SFR measured at different time intervals along the star formation history. Measuring the relative dispersion in the population of galaxies at each time step will reveal an epoch at which both star formation estimates are most closely related.}
    \label{fig:SFR_vs_L1.4GHz_schematic}
\end{figure}

The key aim of this section is to investigate the timescale dependence of the relation between star formation produced by stellar and dust-related processes and that of the subsequent synchrotron radio emission. Figure\,\ref{fig:SFR_vs_L1.4GHz_schematic} shows a schematic of the approach taken in this work to derive the point in time when the two measures of star formation are most closely related. The top panel illustrates the possible star formation histories of two galaxies evolving in complete contrast: either ramping up (blue) or declining (red) in star formation. Also shown as a horizontal dashed line is the SFR that might be measured from the 1.4\,GHz radio continuum. As is likely the case, the radio continuum does not pertain to the SFR now, but rather to some previous epoch of star formation that formed the stars that underwent supernovae and propagated cosmic rays over timescales of several 100\,million years \citep{Condon1992}. The middle row of Figure\,\ref{fig:SFR_vs_L1.4GHz_schematic} shows how the scatter in the \SFRt-\Lradio relation might vary with time if this difference in timescale plays an important role in the apparent relation between SFR estimates. For instance, galaxies with an increasing SFH would be situated above the $\mathrm{SFR}(t=0)$-\Lradio relation at $t=0$\,Gyr as their SFR have since increased and vice versa for a declining SFH. By retracing the star formation histories of galaxies and measuring the intrinsic scatter in the \SFRt-\Lradio relation at different epochs (bottom panel), one should find an epoch with the least dispersion when these SFR indicators are probing a common population of stars.\\

\begin{figure*}
    \centering
    \includegraphics[width=0.495\textwidth]{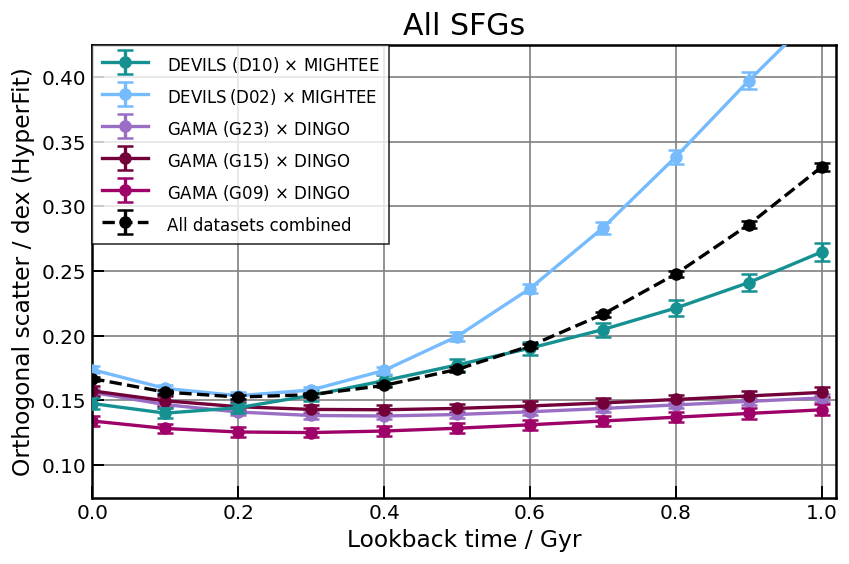}
    \includegraphics[width=0.495\textwidth]{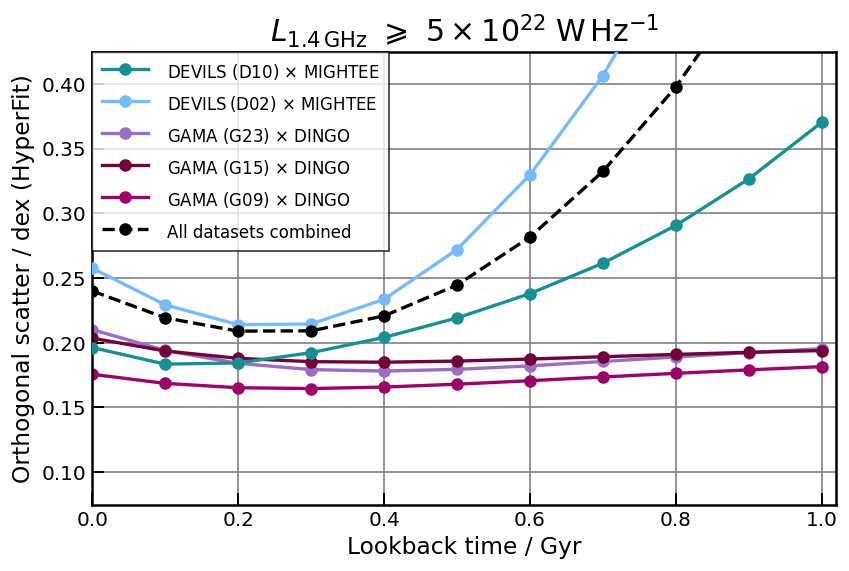}
    \caption{The orthogonal scatter in the $\mathrm{SFR}(t)$--\Lradio relation over lookback time taken from the best-fitting star formation histories modelled by \ProSpect. Each colour corresponds to the different fields used in the analysis, whereas the dashed black line shows the scatter of all samples combined. The right panel shows the variation in scatter only for radio-bright galaxies with 1.4\,GHz luminosities above a threshold of \Lradio$> 5\times10^{22}\,$W\,/Hz. This shows that SFR derived from FUV\,--\,FIR SED models most closely matches the supernova-driven SFR as measured by radio continuum at $\sim$200\,--\,300\,Myr prior in a galaxy star formation histories.}
    \label{fig:scatter_in_SFR_vs_L1.4GHz_over_time}
\end{figure*}

Figure \ref{fig:scatter_in_SFR_vs_L1.4GHz_over_time} shows the orthogonal scatter in the \SFRt-\Lradio relations for star-forming, non-AGN galaxies over 100\,Myr increments in $t_\mathrm{LB}$. The orthogonal scatter was computed as an independent fitting parameter using \Hyperfit, as well as allowing for a free slope and normalisation. Each coloured line represents the scatter over time for the data sets taken individually, whereas the black dashed line shows the orthogonal scatter in the combination of all datasets. For all datasets, the point of minimum scatter is not in the current epoch at $t_\mathrm{LB} = 0$\,Gyr. For the \DEVILSxMIGHTEE datasets, the minimum scatter is found to be at an epoch 100\,--\,300\,Myr prior, with the largest difference seen in the D02 field. The \GAMAxDINGO relations show the least variation between time steps and the point of minimum scatter occurs at a more distant lookback time ($\sim$300\,--\,400\,Myr). However, the lack of significant SFH variations seen in the GAMA fields (e.g. Figure\,\ref{fig:SFH_fractions_vs_lum}) is consistent with no change in scatter over these time intervals.

As highlighted in Figure\,\ref{fig:SFR_vs_L1.4GHz_SFH_subsets}, the largest offset in the \SFRt-\Lradio relation comes from the most star-forming (and hence most luminous) galaxies; i.e., \Lradio$\gtrsim 5\times10^{22}$\,\WperHz. At these luminosities, contamination from AGN is possible, however, such interlopers would shift points downwards on the relation, whereas much of the scatter comes from points above the relation. Below this luminosity, the majority of galaxies exhibit fairly constant slopes in their SFHs --- particularly at the low redshifts covered by the GAMA fields. In the right panel of Figure \ref{fig:scatter_in_SFR_vs_L1.4GHz_over_time}, we show the results of fitting the power-law relation only for galaxies with \Lradio$> 5\times10^{22}$\,\WperHz. The orthogonal scatter is higher above this luminosity cut, however, this also shows a greater reduction in scatter between the \SFRt-\Lradio relations. Combining the datasets (black dashed line) results in a net decrease of 0.031\,dex between the current epoch and $t_\mathrm{LB} = 300$\,Myr, the point of minimum scatter. This implies that the star formation rate as measured by the radio continuum and that derived from the UV\,--\,IR photometry is most closely related at a time 300\,Myr previously.

\section{Discussion}
\label{sec:discussion}
In this section, we discuss the implications of the discrepancy seen between the SED-derived and radio-continuum SFRs and possible interpretations for these results. \citet{Petter2020} suggest that the difference in timescales probed by the emission mechanisms could --- in combination with several other factors --- explain why galaxies with a younger mean stellar age (potentially star bursting) are observed to have a higher infrared luminosity compared to their radio luminosity. Simply not enough time has passed since the starburst for the corresponding synchrotron emission to be detected.

Recently, \citet{Arango-Toro2023} performed a similar analysis using a mass-complete sample of SFGs observed with the VLA at 3\,GHz with SED models fit using \texttt{CIGALE} \citep{Boquien2019}, which includes the non-parametric implementation of star formation histories from \citep{Ciesla2023}. The authors also find a trend in the offset between SED-derived SFRs and those from the radio continuum, however, primarily driven by galaxies with a declining SFH. This study also finds that both SFR indicators converge when reverting to an earlier point in their SFH, specifically by averaging over a period later than 150\,--\,300\,Myr. The results of \citep{Arango-Toro2023} complement the findings of this work, where the greatest offset is caused by an overestimate of the SED-derived SFR from galaxies with a recent increase in their SFH. The implementation of a non-parametric SFH likely explains their ability to capture precipitous decreases in SFH, which can be difficult to model for old galaxies with the truncated skewed-normal SFH (see Section\,\ref{sec:SFH_limitations}).

The origin of the infrared-radio correlation often assumes that synchrotron emission dominates over inverse-Compton emission and that the rate at which supernovae fade is longer than the electron cooling time \citep{Bressan2002, Ivison2010}. \citet{Galvin2016} found that the \qTIR parameter increases in value and scatter with increases in the fraction of thermal (free-free) emission with respect to non-thermal (synchrotron) emission. They suggest that this could be due to the differing timescales, whereby galaxies with more recent starbursts may have a higher thermal fraction and, thus, higher \qTIR due to the delay in producing non-thermal emission from accelerated electrons. A similar result was found by \citet{An2021} who, using 0.33\,--\,3GHz radio continuum observations from MeerKAT and VLA observations, showed that flatter spectral indices resulted in underestimating the \qTIR parameter. More recently, \citet{An2023} used a sample of SFGs detected in both LOFAR (150\,MHz) and GMRT (610\,MHz) to show that, on average, radio spectral indices steepen slightly towards increasing stellar mass. The authors suggest that spectral ageing due to the energy loss of CRes and thermal free-free absorption are possible physical mechanisms that drive this correlation towards higher masses.

\citet{Roussel2003} provide an alternative explanation for an SFR offset that could be explained by a difference in the initial mass function between low and high-mass galaxies, which relies on the fact that the mass spectrum of stars capable of producing UV\,--\,FIR emission extends to lower masses than those that will undergo supernovae. A higher ultraviolet/infrared luminosity for a given radio luminosity could be achieved if the IMF had a very steep slope, causing there to be fewer high-mass stars. However, it is difficult to reconcile this with the fact that a better agreement is achieved between \Lradio and SFR$(t=200\,\mathrm{Myr})$ than with SFR$(t=0\,\mathrm{Myr})$. Furthermore, without a clear understanding of the intrinsic variation of the IMF between galaxies, it is difficult to discern the relative importance.

\subsection{Investigating the impact of the SFH slope on the SFR-\texorpdfstring{\Lradio}{L1.4GHz} relation}
\label{sec:discussion.radio_SFR_calibration}
\begin{figure}
    \centering
    \includegraphics[width=\columnwidth]{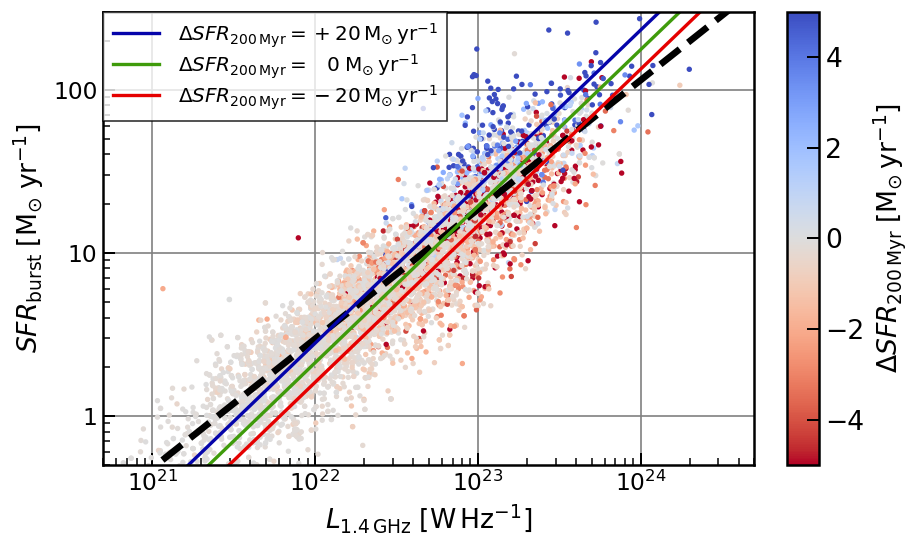}
    \caption{The SFR against the 1.4\,GHz radio luminosity coloured by \DeltaSFRMyr for all datasets combined. The black dashed line is the resulting best-fitting relation for the two-dimensional \LTIR-\Lradio relation. The coloured solid lines represent the hyperplane fit to the SFR-\Lradio relation with \DeltaSFRMyr as the additional dimension when fit with \Hyperfit, intersecting Equation\,\ref{eqn:SFR_hyperplane} at the values for a constant SFH slope and \DeltaSFRMyr$=\pm$20\,\MSunperyear.}
    \label{fig:SFR_vs_L1.4GHz_hyperplane}
\end{figure}

As has been shown in the previous sections, the 1.4\,GHz radio continuum emission of a galaxy can be higher or lower than what is measured from the recent SFR (as probed by UV\,--\,IR emission) depending on the trajectory of that galaxy's recent star formation history. In this section, we attempt to account for the variation in the recent SFH slope by fitting a hyperplane to the SFR-\Lradio relation with the inclusion of the \DeltaSFRMyr quantity as an additional axis. We again use \Hyperfit to optimise this best-fitting plane, which gives the following relation:

\begin{equation}
    \begin{split}
    \log_{10}\left(\frac{SFR}{\mathrm{M_{\odot}\,yr^{-1}}}\right) = (0.96\pm0.01)\log_{10}\left(\frac{L_\mathrm{1.4\,GHz}}{5\times10^{22}\,\mathrm{W\,Hz^{-1}}}\right) \\ 
    +\;(0.006\pm0.001)\left(\frac{\Delta SFR_\mathrm{200\,Myr}}{\mathrm{M_{\odot}\,yr^{-1}}}\right) + (0.996\pm0.003),
    \end{split}
    \label{eqn:SFR_hyperplane}
\end{equation}

\noindent
with an orthogonal scatter of $\sigma_{\perp} = 0.152\pm0.002$\,dex --- marginally lower than for the relation without \DeltaSFRMyr. Figure\,\ref{fig:SFR_vs_L1.4GHz_hyperplane} shows the SFR-\Lradio relation for the combined dataset with overplotted lines intersecting the hyperplane expression of Equation\,\ref{eqn:SFR_hyperplane} at the values of $\pm$20\,\MSunperyear. The best-fitting slope with \DeltaSFRMyr of 0.006\,dex implies that the impact of the distribution of star formation histories in estimating the SFR from a \Lradio measurement may be relatively minor. For instance, for the SFR of a galaxy to be overestimated by 0.1\,dex ($\sim26$\,\%), a galaxy would have to be experiencing a change in its SFR of $\sim$17\,\MSunperyear over the past 200\,Myr. Such a rapid change in SFR is uncommon in the local Universe, suggesting SFR timescales only have a minor impact on local radio continuum calibrations of SFR. However, this is not likely to be the case for galaxies observed at earlier epochs of the Universe.\\

We have shown that the slope of the power law relation describing the IRRC is not unity, thus calibrations of SFR from 1.4\,GHz that do not account for this underlying luminosity dependence \citep[e.g.][]{Yun2001, Bell2003} will inherently over-estimate SFR at brighter radio luminosities. For instance, it has been shown in \citet{Molnar2021} that these SFR calibrations could reach an excess of $\sim0.2$\,dex compared to SED-derived SFRs of \citep{Salim2016} at \Lradio$> 10^{23}$\,\WperHz. Forcing \qTIR to be a constant value requires introducing a piece-wise expression, such as the \citet{Bell2003} and \citet{Hopkins2003} prescriptions, to account for this luminosity dependence. Recent studies have attempted to account for the diminishing infrared \citep{Lacki2010b} by incorporating ultraviolet photometry into their proxies for star formation rate \citep{Davies2017, Delvecchio2021}. As we have shown in Figure\,\ref{fig:SFR_vs_L1.4GHz}, incorporating the ultraviolet emission --- tracing OB stars on timescales of $\sim20$\,--\,$80$\,Myr --- likely dissociates the radio continuum emission further from the FUV\,--\,FIR. Correcting for this effect by tracing the star formation histories back to a previous epoch has the effect of linearising the SFR-\Lradio relation, however, the non-unity slope remains. This dependence on \Lradio must be accounted for when calibrating SFRs based on the radio luminosity alone.

\subsection{Limitations of the skewed log-normal SFH}
\label{sec:SFH_limitations}
In detail, the SED-constrained SFHs are unable to resolve variations in star formation on timescales shorter than 100\,Myr and the smooth skewed Gaussian profile itself does not explicitly model starbursts. This means our findings are only sensitive to the smooth overall changes suppressing or enhancing star formation in galaxies. A finer time resolution may more precisely isolate the average time delay between measuring commensurate star formation rates in radio and ultraviolet\,--\,infrared wavelengths. However, a direct link cannot be made without knowing the exact stochastic history of a galaxy's evolution --- currently only achievable in semi-analytic models and hydrodynamical simulations.

To assess the impact of using this simplified function, we have compared the true star formation histories of simulated galaxies from the semi-analytic model \texttt{Shark} \citep{Lagos2018} with the skewed log-normal function reproduced with \ProSpect fits on mock photometric measurements of the simulated SED from \citet{Bravo2022}. We find a general agreement between SFHs, however, the skewed log-normal function tends to miss recent bursts of star formation, particularly in high-mass galaxies where the \ProSpect model is influenced by the bulk of stars being formed at much earlier times. The impact of these modelled SFHs will be explored in greater detail in Davies et al. (submitted, MNRAS) and progress has already been made in implementing separate SFHs for the stellar populations of bulges and disks \citep{Robotham2022, Bellstedt2024}, which allows for a greater variety of realistic SFHs to be modelled.

    
    

%

\section{Conclusions}
\label{sec:conclusions}
In this paper, we investigate the infrared-radio correlation (IRRC) for $\sim$5,500 star-forming galaxies (excluding contamination from AGN) by combining multi-wavelength datasets across five fields from both the DEVILS and GAMA surveys with corresponding 1.4\,GHz radio continuum detections from the MIGHTEE and DINGO surveys. In addition to the recent SFR and infrared luminosity derived from \ProSpect, we also measure the variation in the star formation history over the last 200\,Myr (\DeltaSFRMyr) to explore potential causes for the non-unity slope in the IRRC and the implications this has for estimating star formation rates from radio luminosities. We summarise our findings as follows:

\begin{itemize}
    \item The combination of radio continuum observations from the SKAO precursor telescopes of MeerKAT and ASKAP with multi-wavelength datasets reproduces the well-known IRRC with a tight scatter of 0.14\,dex and sub-unity power law slope of 0.921$\pm$0.004. We see that the logarithmic ratio of the infrared-to-radio (\qTIR) exhibits little-to-no trend with redshift, however, a slight dependence on stellar mass is observed with \qTIR decreasing by 0.08\,dex per 1\,dex increase in stellar mass (see Section\,\ref{sec:qTIR_by_redshift}).

    \item A similarly tight relation with scatter of 0.16\,dex is found when replacing the infrared luminosity (itself a proxy for prolonged star formation) with the SED-derived SFR (see Section\,\ref{sec:SFR_vs_L1.4GHz}). The resulting slope of 0.87 remains sub-unity, suggesting that the non-linear scaling of \qTIR is implicit in the correlation between \Lradio and SFR.

    \item As has been seen in previous studies, a break in the SFR-\Lradio relation is seen that demands a steeper slope towards higher \Lradio. It is also at high SFR (and hence, \Lradio) that we observe a trend with the star formation histories of galaxies, whereby galaxies with a \DeltaSFRMyr $>$ 0.5\,\MSunperyear are offset to higher SED-derived SFRs compared to their \Lradio. This is also evidenced by the observation that the \qTIR of galaxies with an increasing SFH is 0.1\,dex larger on average than the population of constant SFHs.
    
    \item We show that backtracking the SFR of all galaxies along their SFH will both linearise the SFR-\Lradio relation and reduce the overall scatter. The minimum scatter in the SFR(t)-\Lradio is reached at a point 200\,--\,300\,Myr prior in their SFHs (see Section\,\ref{sec:SFR_vs_L1.4GHz_over_time}). This is consistent with theoretical predictions of the timescales required to accelerate and disperse the cosmic ray electrons that produce synchrotron emission in SFGs.
    
    \item We explore the \DeltaSFRMyr on the SFR-\Lradio relation by incorporating the SFR slope as an additional axis for fitting as a hyperplane. The impact of this parameter is small in the nearby Universe where the majority of galaxies have relatively constant SFHs. The orthogonal scatter in the best-fitting relation is therefore only marginally ($-0.01$\,dex) reduced when including \DeltaSFRMyr.
\end{itemize}

Radio emission borne out of star formation processes will constitute the dominant population at low luminosities in upcoming large radio continuum surveys, such as those being conducted with the SKAO and ngVLA. Furthermore, the sensitivity of the full SKAO will allow for detecting substantial numbers of SFGs out to redshifts $z \lesssim 6$ \citep{Jarvis2015, Bonaldi2019}, where galaxies are experiencing more rapid fluctuations in their star formation.

Therefore, incorporating the impact of SFR timescales when calibrating SFRs from radio continuum images will become increasingly important as it becomes a de facto standard in modern studies of galaxy evolution.

\section*{Acknowledgements}
We express our gratitude to the referee for their insightful comments and constructive feedback, which significantly improved the quality of this paper.\\

DEVILS is an Australian project based around a spectroscopic campaign using the Anglo-Australian Telescope. The DEVILS input catalogue is generated from data taken as part of the ESO VISTA-VIDEO \citep{Jarvis2013} and UltraVISTA \citep{McCracken2012} surveys. DEVILS is part funded via Discovery Programs by the Australian Research Council and the participating institutions. The DEVILS website is \url{https://devilsurvey.org.} The DEVILS data is hosted and provided by AAO Data Central (\url{https://datacentral.org.au/}).

GAMA is a joint European-Australasian project based around a spectroscopic campaign using the Anglo-Australian Telescope. The GAMA input catalogue is based on data taken from the Sloan Digital Sky Survey and the UKIRT Infrared Deep Sky Survey. Complementary imaging of the GAMA regions is being obtained by a number of independent survey programmes including GALEX MIS, VST KiDS, VISTA VIKING, WISE, Herschel-ATLAS, GMRT and ASKAP providing UV to radio coverage. GAMA is funded by the STFC (UK), the ARC (Australia), the AAO, and the participating institutions. The GAMA website is \url{http://www.gama-survey.org/}.

The MeerKAT telescope is operated by the South African Radio Astronomy Observatory, which is a facility of the National Research Foundation, an agency of the Department of Science and Innovation. We acknowledge the use of the ilifu cloud computing facility --- www.ilifu.ac.za, a partnership between the University of Cape Town, the University of the Western Cape, Stellenbosch University, Sol Plaatje University and the Cape Peninsula University of Technology. The Ilifu facility is supported by contributions from the Inter-University Institute for Data Intensive Astronomy (IDIA --- a partnership between the University of Cape Town, the University of Pretoria and the University of the Western Cape), the Computational Biology division at UCT and the Data Intensive Research Initiative of South Africa (DIRISA).

The Australian SKA Pathfinder is part of the Australia Telescope National Facility which is managed by CSIRO. Operation of ASKAP is funded by the Australian Government with support from the National Collaborative Research Infrastructure Strategy. ASKAP uses the resources of the Pawsey Supercomputing Centre. Establishment of ASKAP, the Murchison Radio-astronomy Observatory and the Pawsey Supercomputing Centre are initiatives of the Australian Government, with support from the Government of Western Australia and the Science and Industry Endowment Fund. We acknowledge the Wajarri Yamatji people as the traditional owners of the Observatory site

This work is based on data products from observations made with ESO Telescopes at the La Silla Paranal Observatory under ESO programme ID 179.A-2005 (Ultra-VISTA) and ID 179.A- 2006 (VIDEO) and on data products produced by CALET and the Cambridge Astronomy Survey Unit on behalf of the Ultra-VISTA and VIDEO consortia. Based on observations obtained with MegaPrime/MegaCam, a joint project of CFHT and CEA/IRFU, at the Canada-France-Hawaii Telescope (CFHT) which is operated by the National Research Council (NRC) of Canada, the Institut National des Science de l’Univers of the Centre National de la Recherche Scientifique (CNRS) of France, and the University of Hawaii. This work is based in part on data products produced at Terapix available at the Canadian Astronomy Data Centre as part of the Canada-France-Hawaii Telescope Legacy Survey, a collaborative project of NRC and CNRS. The Hyper Suprime-Cam (HSC) collaboration includes the astronomical communities of Japan and Taiwan, and Princeton University. The HSC instrumentation and software were developed by the National Astronomical Observatory of Japan (NAOJ), the Kavli Institute for the Physics and Mathematics of the Universe (Kavli IPMU), the University of Tokyo, the High Energy Accelerator Research Organization (KEK), the Academia Sinica Institute for Astronomy and Astrophysics in Taiwan (ASIAA), and Princeton University. Funding was contributed by the FIRST program from Japanese Cabinet Office, the Ministry of Education, Culture, Sports, Science and Technology (MEXT), the Japan Society for the Promotion of Science (JSPS), Japan Science and Technology Agency (JST), the Toray Science Foundation, NAOJ, Kavli IPMU, KEK, ASIAA, and Princeton University.

MJJ, IHW and CLH acknowledge generous support from the Hintze Family Charitable Foundation through the Oxford Hintze Centre for Astrophysical Surveys. MJJ acknowledges the support of the STFC consolidated grant [ST/S000488/1] and [ST/W000903/1] and from a UKRI Frontiers Research Grant [EP/X026639/1]. CLH acknowledges support from the Leverhulme Trust through an Early Career Fellowship. ID acknowledges support from INAF Minigrant "Harnessing the power of VLBA towards a census of AGN and star formation at high redshift".

This work made use of \texttt{python}, specifically the following packages: \texttt{astropy}:\footnote{http://www.astropy.org}, \citep{AstropyCollaboration2022}; \texttt{matplotlib} \citep{Hunter2007}; \texttt{numpy} \citep{vanderWalt2011, Harris2020}; \texttt{scipy} \citep{Virtanen2020}.

\section*{Data Availability}
The DEVILS and GAMA data products used throughout this paper are currently available for use by internal team members conducting proprietary science with public access being made available in upcoming data releases.\\
The MIGHTEE Early Science data used for this work is discussed in depth in \citep{Heywood2022}. The cross-matched catalogue is described in the work of \citet{Whittam2024} and the classification of radio galaxies into SFGs and AGN is described in \citet{Whittam2022}. The catalogues will be released in accompaniment with their work. The derived data produced in this work can be found in the article and supplementary material, or can be shared upon reasonable request to the corresponding author.



\bibliographystyle{mnras}
\bibliography{references} 



\appendix

\section{Distribution of qTIR for AGN}
\label{appndx:qTIR_distribution}
\begin{figure}
    \centering
    \includegraphics[width=\columnwidth]{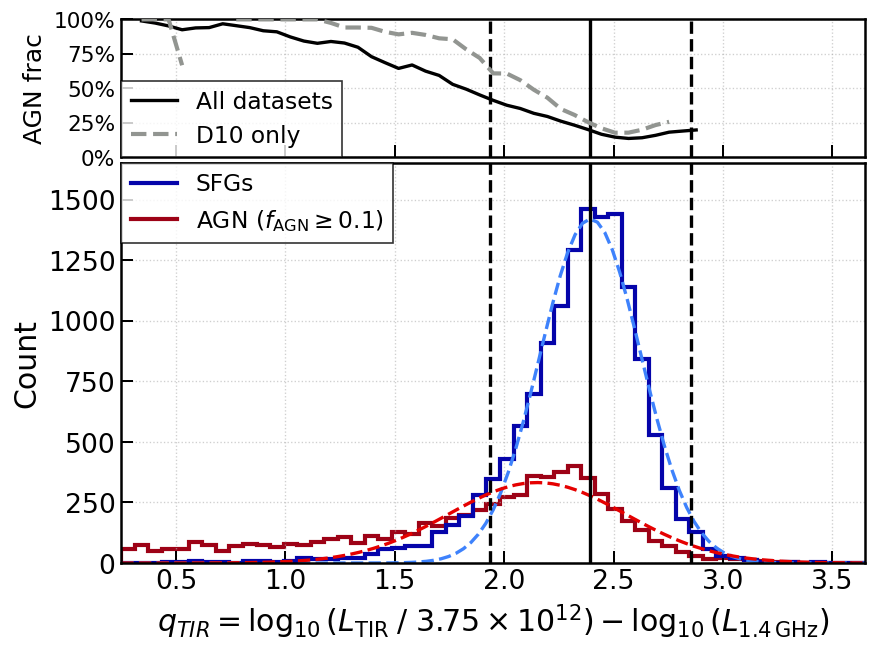}
    \caption{Bottom panel shows the distributions of \qTIR values for the sample of SFGs (blue) and AGN (red) selected via the \fAGN cut. Overplotted are the best-fitting Gaussian profiles for these distributions with the black vertical lines showing the mean value, \aveqTIR (solid), and $\pm2\,\sigma$ (dashed). The top panel shows the fraction of AGN galaxies within a given \qTIR bin smoothed by a running mean across adjacent data points for combined samples (black) and the D10 sample only (grey dashed). At 2$\sigma$ below the mean \qTIR, almost all galaxies show some signature of AGN in the D10 sample, which includes additional AGN criteria in X-ray, optical, mid-IR and radio wavelengths.}
    \label{fig:qTIR_hists}
\end{figure}

Figure\,\ref{fig:qTIR_hists} shows the distribution of \qTIR for star-forming galaxies (blue) and those identified as AGN (red) either by having an \fAGN$>0.1$ or, in the central 0.8\,deg$^{2}$ of the D10 region, being flagged as AGN in any one of the X-ray, optical, mid-infrared, broad- and narrow-band emission line metrics described in \citet{Whittam2022}. The D10 sample thus provides a cleaner sample of purely star-forming galaxies with the least contamination from AGN not detected in the infrared SED templates of \citet{Thorne2022b}. Note that no star formation activity cut has been imposed on the AGN sample, thus some of these sources would likely be selected in a sample of star-forming galaxies.

The population of all SFGs appears to be well-modelled by a Gaussian, centred on \aveqTIR = 2.4 with a dispersion of 0.21\,dex. As expected, the AGN sample is offset to lower values with average \aveqTIR = 2.15, spread over a larger range of 0.40\,dex that skews heavily towards lower \qTIR. Although AGN-selected galaxies have on average lower \qTIR values, the overlap in these samples is significant in their \qTIR distributions, meaning that no one value in \qTIR can safely separate SFGs from AGN without some contamination. The top panel shows this contamination of AGN sources as a function of \qTIR for the combined samples and the COSMOS region only (dashed line). At 2$\sigma$ below the peak \qTIR, $\sim$50\% of galaxies are identified as having an AGN (almost all galaxies in COSMOS). However, above 2$\sigma$, there remains a large fraction of galaxies with a significant AGN contribution. This result demonstrates that using a cut in the \qTIR distribution \emph{alone} is not sufficient to provide a sample free from significant AGN contamination, particularly LERGs.

\bsp	
\label{lastpage}
\end{document}